\documentclass[aps,preprint]{revtex4}%
\usepackage{amsfonts}
\usepackage{amsmath}
\usepackage{amssymb}
\usepackage{graphicx}%
\providecommand{\U}[1]{\protect\rule{.1in}{.1in}}

\begin{document}
\preprint{ }
\title[Dark sector ]{Macroscopic theory of dark sector}
\author{Boris E. Meierovich}
\affiliation{P. L. Kapitza institute for physical problems}
\keywords{Dark enerdy, dark matter, vector fields, galaxy rotation curves, universe evolution}
\begin{abstract}
A simple Lagrangian with squared covariant divergence of a vector field as a
kinetic term turned out an adequate tool for macroscopic description of the
dark sector. The zero-mass field acts as the dark energy. Its energy-momentum
tensor is a simple additive to the cosmological constant. Massive fields
$\phi_{I}$ with $\phi^{K}\phi_{K}<0$ and $\phi^{K}\phi_{K}>0$ describe two
different forms of dark matter. The space-like $\left(  \phi^{K}\phi
_{K}<0\right)  $ massive vector field is attractive. It is responsible for the
observed plateau in galaxy rotation curves. The time-like $\left(  \phi
^{K}\phi_{K}>0\right)  $ massive field displays repulsive elasticity. In
balance with dark energy and ordinary matter it provides a four parametric
diversity of regular solutions of the Einstein equations describing different
possible cosmological and oscillating non-singular scenarios of evolution of
the universe. In particular, the singular big bang turns into a regular
inflation-like transition from contraction to expansion with the accelerate
expansion at late times. The fine-tuned Friedman-Robertson-Walker singular
solution is a particular limiting case at the lower boundary of existence of
regular oscillating solutions in the absence of vector fields. The simplicity
of the general covariant expression for the energy-momentum tensor allows to
display the main properties of the dark sector analytically. Although the
physical nature of dark sector is still unknown, the macroscopic theory can
help analyzing the role of dark matter in astrophysical phenomena without
resorting to artificial model assumptions.

\end{abstract}
\maketitle
\tableofcontents

\section{\label{Introduction}Introduction}

Currently there are two most intriguing long standing problems in astrophysics
pointing to the existence of so called "hidden sector", containing "dark
energy" and "dark matter". So far their interaction with the ordinary matter
(baryons and leptons) is observed only via gravitation.

The first problem, named "galaxy rotation curves", appeared in 1924, after J.
H. Oort discovered the galactic halo, a group of stars orbiting the Milky Way
outside the main disk \cite{Oort 1924}. In 1933, F. Zwicky \cite{Zwicky 1933}
postulated "missing mass" to account for the orbital velocities of galaxies in clusters.

The second problem is the accelerated expansion of the universe discovered
through observations of distant supernovae by Adam G. Riess, Brian P. Schmidt,
Saul Perlmutter, and their colleagues in 1998 \cite{Adam Riess}%
,\cite{Perlmutter}.

At first glance, these two problems have little to do with one another. The
accelerated expansion of the universe indicates the existence of a hidden
mechanism of repulsion \cite{Meierovich1}, while the plateau of the galaxy
rotation curves is the result of additional attraction caused by the dark
matter \cite{Meierovich}. Macroscopic approach to the dark sector problems,
based on the analysis of vector fields in general relativity, provides an
appropriate universal tool for the theoretical description of both these
phenomena. The spacelike massive vector field is attractive. It is responsible
for the observed plateau in galaxy rotation curves. The timelike massive
vector field displays repulsive elasticity. In the scale of the whole universe
it is the source of accelerated expansion. Naturally, the previous solutions
of the Einstein equations, describing the expansion of the universe filled
with the mutually attracting matter only, inevitably contained a singularity.
Inclusion of the repulsive dark matter into consideration allows the existence
of nonsingular solutions describing various possible regular scenarios of
evolution of the universe.

This article contains the macroscopic theory of dark sector, based on the
analysis of vector fields in general relativity. The step by step derivations
are accompanied by the references to the benchmark achievements of the
predecessors. The main attention is paid to clarify the validity of basing assumptions.

Vector fields are used to describe quantum particles of the ordinary matter
\cite{Landau-Lifshits4}. A zero-mass particle -- photon -- is a quantum of
electromagnetic field obeying Maxwell equations. Massive bosons obey Proca
equations \cite{Proca}. The relation of spinors and vectors
(\cite{Landau-Lifshits4}, page\ 88) facilitates establishing Dirac equations
for fermions.

Field equations for quantum particles are easily established in accordance
with the properties of their free motion in plane geometry. If necessary (for
the secondary quantization, for instance), the Lagrangian of a particle is
then constructed in such a way, that the field equations minimize the
functional of action. This approach is convenient for description of already
known particles. However, it doesn't help to describe the unknown substance of
the dark sector.

In general relativity, the standard approach starting from a general form of
the Lagrangian of a vector field is capable to describe not only the already
known particles. It is reasonable to start from a general form of the
Lagrangian of a vector field in general relativity, and derive the vector
field equations and the energy-momentum tensor. Excluding the terms associated
with the ordinary matter, one can separate the Lagrangian having a chance to
describe the dark sector. The separation of the Lagrangian of the dark sector
is necessary, especially if the ordinary matter in the universe is considered
as a continuous medium with the macroscopic energy-momentum tensor
(\ref{Tom_IK}).\ Otherwise the ordinary matter would be taken into account
twice: as a medium with the energy-momentum tensor (\ref{Tom_IK}), and as
quantum particles described by the vector field.

It turns out that the most simple Lagrangian of a vector field (\ref{L dark})
(with the squared covariant divergence as a kinetic term) allows to describe
the main observed manifestations of the dark sector completely within the
frames of the minimal general relativity. In this case, the massless field
corresponds to the dark energy, the massive spacelike field $\left(  \phi
^{K}\phi_{K}<0\right)  $ is \ responsible for a plateau in the galaxy rotation
curves, and the massive timelike vector field $\left(  \phi^{K}\phi
_{K}>0\right)  $ displays the repulsive elasticity. The competition of
repulsive dark matter and attractive ordinary matter leads to a variety of
possible regular scenarios of evolution of the universe.

\begin{figure}
[ptbh]
\begin{center}
\includegraphics[
trim=0.000000in 0.000000in -0.001109in 0.004203in,
height=7.2137cm,
width=9.5092cm
]%
{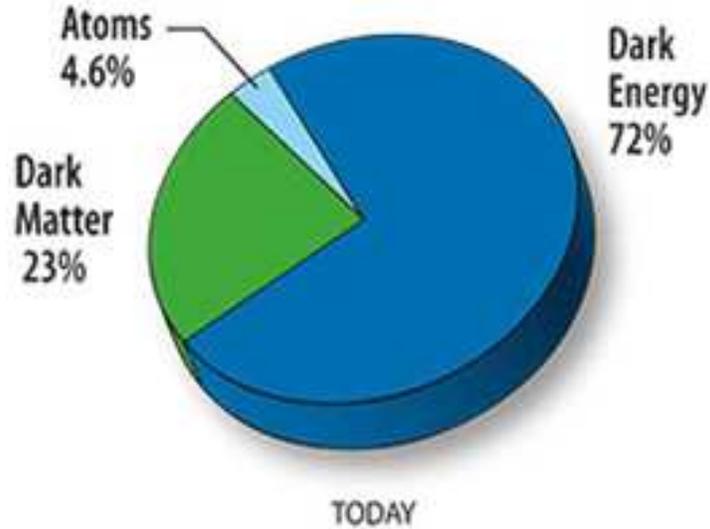}%
\caption{Staff of the universe \cite{NASA sliced cake}}%
\label{fig1}%
\end{center}
\end{figure}

According to the NASA \textquotedblleft sliced cake\textquotedblright\ diagram
\cite{NASA sliced cake}, see Figure \ref{fig1},%
today there is only 4.6\% of the ordinary matter among the staff of the
universe. All other 95\% is the unknown substance, referred to as the dark
matter and dark energy. For this reason it is natural to start with the
analysis of the properties of dark sector, and add the ordinary matter into
consideration after the role of dark sector is clarified.

The main properties of vector fields in general relativity are recalled in
Section \ref{V f in G R} in order to clarify the specifics of the approach
based on the principle of regularity. The features of the Proca equations
allow to separate the terms in the Lagrangian, which are not connected with
the ordinary matter. It turns out that a simple Lagrangian (with the squared
covariant divergence as a kinetic term) is a proper tool for macroscopic
description of main observed properties of the dark sector. The necessary
conditions of regularity for a spacelike and a timelike vectors are different.
Nevertheless, the field equations and the energy-momentum tensor have the same
covariant form for both kinds of vector fields. The simplicity of equations
allows to get analytical solutions in the most interesting cases. The galaxy
rotation curves, driven by the spacelike vector fields, are derived in Section
\ref{Rotation curves}. Various uniform and isotropic scenarios of evolution of
the universe under the joint influence of the zero-mass vector field, the
timelike massive vector field, and the ordinary matter, are analyzed in
Section \ref{Cosmology}. The current situation with the dark sector is
summarized in Section \ref{Summary}. Some major astrophysical problems are
specified, where the macroscopic theory can be applied, helping to avoid the
unnecessary model assumptions. Unraveling the still unknown physical nature of
the dark energy and dark matter remains the most pressing issue.

\section{\label{V f in G R}Vector fields in general relativity}

\subsection{\label{Lag of vec fie}Lagrangian of a vector field}

Within the frames of minimal general relativity (field equations no higher
than of the second order), the Lagrangian of a vector field $\phi_{I}$ is a
scalar consisting of bilinear combinations of the covariant derivatives
$\phi_{I;K}$ and a scalar potential $V(\phi^{K}\phi_{K})$. A bilinear
combination of the covariant derivatives is a 4-index tensor
\[
S_{IKLM}=\phi_{I;K}\phi_{L;M}.
\]
A general form of the scalar $S$, formed via contractions of $S_{IKLM}$, is
\begin{equation}
S=\left(  ag^{IK}g^{LM}+bg^{IL}g^{KM}+cg^{IM}g^{KL}\right)  S_{IKLM},
\label{Scalar S}%
\end{equation}
where $a,b,$ and $c$ are arbitrary constants. Therefore, a general form of the
Lagrangian $L$ of a vector field $\phi_{I},$
\begin{equation}
L\left(  \phi_{I},\frac{\partial\phi_{I}}{\partial x^{K}},g^{IK}%
,\frac{\partial g_{IK}}{\partial x^{L}}\right)  =a(\phi_{;M}^{M})^{2}%
+b\phi_{;M}^{L}\phi_{L}^{;M}+c\phi_{;M}^{L}\phi_{;L}^{M}-V(\phi_{M}\phi^{M}),
\label{Lagrangian}%
\end{equation}
contains three kinetic terms with the arbitrary coefficients $a,b,$ and $c$.
Applying the least action principle, in view of
\[
\phi^{K}=g^{IK}\phi_{I},\ \ \phi_{I;K}=\frac{\partial\phi_{I}}{\partial x^{K}%
}-\Gamma_{IK}^{L}\phi_{L},\ \ \ \Gamma_{IK}^{L}=\frac{1}{2}g^{LM}\left(
\frac{\partial g_{MI}}{\partial x^{K}}+\frac{\partial g_{MK}}{\partial x^{I}%
}-\frac{\partial g_{IK}}{\partial x^{M}}\right)  ,....\text{ },
\]
it is convenient to consider the Lagrangian (\ref{Lagrangian}) as a function
of $\phi_{I},\frac{\partial\phi_{I}}{\partial x^{K}},g^{IK},\frac{\partial
g_{IK}}{\partial x^{L}}$ as independent variables.

\subsubsection{Bumblebee models}

Stricktly speaking, (\ref{Lagrangian}) is not yet the most general form of the
Lagrangian of a vector field. Scalars can be made out of $S_{IKLM}$ not only
via contractions, but also by convolutions with participation of $\phi^{I}%
,$\ like $g^{IK}\phi^{L}\phi^{M}S_{IKLM},$ $R^{IK}\phi^{L}\phi^{M}S_{IKLM},$
and so on. In principle, the number of independent constants can exceed 3.

The direction of a vector is specified, and solutions can be less symmetric
than the initial Lagrangian. In this case the symmetry of a system is
considered as spontaneously broken. Spontaneous breaking of Loretz symmetry by
a vector field $\phi^{I},$ having a non-zero expectation value in vacuum, is a
subject of so called "bumblebee theories", see \cite{Kostelesky 1}-\cite{Bluhm
et al} and references therein. Within the frames of field equations no higher
than of the second order, the most general form of the action is%
\begin{equation}
S=\int d^{4}x\sqrt{-g}\left[  R+J^{IKLM}R_{IKLM}+K^{IKLM}\phi_{I;K}\phi
_{L;M}-V\left(  g_{IK}\phi^{I}\phi^{K}\right)  \right]  .
\label{Most general Lagrangian}%
\end{equation}
Here $R_{IKLM}$ is the Riemann tensor of curvature, $R=g^{IL}g^{KM}R_{IKLM},$
$J^{IKLM}$\ and $K^{IKLM}$\ are arbitrary tensors formed out of the vector
$\phi_{I}$ and the metric tensor $g_{IK}.$ In practice people restrict
themselves by simplified models. For instance, Seifert \cite{Seifert}
considered recently the particular case $J^{IKLM}=0,$ $K^{IKLM}\phi_{I;K}%
\phi_{L;M}=\alpha\widetilde{g}^{IK}\widetilde{g}^{LM}\left(  \phi_{I;L}%
-\phi_{L;I}\right)  \left(  \phi_{K;M}-\phi_{M;K}\right)  ,$ where
$\widetilde{g}^{IK}=g^{IK}+\beta\phi^{I}\phi^{K},$ and $\alpha$ and $\beta$
ars constants. The potential $V\left(  \phi^{L}\phi_{L}\right)  $ is taken to
have a minimum at some non-zero value of its argument $\phi^{L}\phi_{L}.$ In
this model, a perturbation of a Lorentz-violating vector field can be
interpreted as a photon field.

Why "bumblebee"? Perhaps, there was something looking as strange as the
ability of the insect bumblebee to fly successfully despite of being sometimes
questioned on theoretical grounds \cite{Dickinson}. By the way, a possibility
of existence of macroscopic objects, moving at the speed of light and having
zero mass due to gravitational mass-defect, had been mentioned by Andreev back
in 1973 \cite{Andreev},\cite{Andreev 2}.

While a vector $\phi_{I}$ remains small compared to its vacuum expectation
value at a minimum of $V\left(  \phi^{K}\phi_{K}\right)  ,$ the Lagrangian
(\ref{Lagrangian}) is sufficient. However, in case of a small mass
(\ref{m << H}) the longitudinal massive field $\phi_{0}\sim m^{-1}$, see
(\ref{fi(z) = mu>>1}). If in the process of compression the field comes close
to its vacuum expectation value, then a phase transition with spontaneous
symmetry breaking occurs. The bumblebee approach looks promising for future
consideration of a symmetry breaking phase transition in the state of maximum compression.

\subsubsection{Specificity of curved space-time}

The third kinetic term $c\phi_{;M}^{L}\phi_{;L}^{M}$ in (\ref{Lagrangian}) can
be transformed via differentiation by parts to
\[
\phi_{;M}^{L}\phi_{;L}^{M}=\phi_{;L}^{L}\phi_{;M}^{M}+\left(  \phi_{;L}%
^{M}\phi^{L}-\phi_{;L}^{L}\phi^{M}\right)  _{;M}+\left(  \phi_{;L;M}^{L}%
-\phi_{;M;L}^{L}\right)  \phi^{M}.
\]
Here $\left(  \phi_{;L}^{M}\phi^{L}-\phi_{;L}^{L}\phi^{M}\right)  _{;M}$ is a
total differential. It does not change the integral of action, and one can use
the equivalent Lagrangian%
\[
L=\left(  a+c\right)  (\phi_{;M}^{M})^{2}+b\phi_{;M}^{L}\phi_{L}^{;M}+c\left(
\phi_{;L;M}^{L}-\phi_{;M;L}^{L}\right)  \phi^{M}-V(\phi_{M}\phi^{M})
\]
instead of (\ref{Lagrangian}).

In General Relativity the second covariant derivative of a vector is not
invariant against the replacement of the order of differentiation:
\begin{equation}
\phi_{;L;M}^{I}-\phi_{;M;L}^{I}=R_{\text{ }KML}^{I}\phi^{K}.
\label{Changing order of differ}%
\end{equation}
$R_{IKLM}$ is the Riemann tensor of curvature. Hence%

\begin{equation}
\phi_{;L;M}^{L}-\phi_{;M;L}^{L}=R_{\text{ }KML}^{L}\phi^{K}=R_{KM}\phi^{K},
\label{Difference between A and B}%
\end{equation}
where $R_{KM}$ is the Ricci tensor. In a curved space-time $R_{KM}\neq0,$ and
the term $c\left(  \phi_{;L;M}^{L}-\phi_{;M;L}^{L}\right)  \phi^{M}$ affects
the integral of action. From the point of view of general relativity all three
kinetic terms in (\ref{Lagrangian}) are equally important.\ If we adhere the
view that in the quantum physics each elementary particle is a quantum of some
field, and vice versa, each field corresponds to its own quantum particle
\cite{Rubakov}, then, in principle, any linear combination of the three
kinetic terms in the Lagrangian (\ref{Lagrangian}) could be associated with
some sort of matter.

In the plane geometry $R_{KM}=0,$ the term $c\left(  \phi_{;L;M}^{L}%
-\phi_{;M;L}^{L}\right)  \phi^{M}$ \ drops out, a covariant derivative
$\phi_{;M}^{L}$\ reduces to the ordinary one $\phi_{,M}^{L}$, and there are
only two arbitrary constants ($b,$ and $\widetilde{c}=a+c$) in the Lagrangian:%
\[
L_{\text{plane}}=b\phi_{,M}^{L}\phi_{L}^{,M}+\widetilde{c}\phi_{,M}^{L}%
\phi_{,L}^{M}-V(\phi_{M}\phi^{M}),\qquad R_{KM}=0.
\]

It is convenient to classify the vector fields $\phi_{I}$ according to their
properties of invariance and symmetry.

The sign of the scalar $\phi_{K}\phi^{K}$ is invariant against the arbitrary
transformations of coordinates. Therefore, if there is no interaction other
than via gravitation, there can be three different independent vector fields
with $\phi_{K}\phi^{K}<0,$ $\phi_{K}\phi^{K}=0,$ $\phi_{K}\phi^{K}>0.$ If
$\phi_{K}\phi^{K}\neq0,$ then in general relativity one can choose a reference
frame where either $\phi_{0}=0$ when $\phi_{K}\phi^{K}<0$ (spacelike
vector)$,$ or $\phi_{I>0}=0$ if $\phi_{K}\phi^{K}>0$ (timelike vector)$.$
$\phi_{K}\phi^{K}=0$ is a separate case. The field equations for an ordinary
massive particle are easily derived basing on the statement that in the plain
space-time there is a reference frame where the particle is at rest
\cite{Landau-Lifshits4}. Hence, the ordinary massive particles are described
by the spacelike fields. For the zero-mass particles (such as photons) there
is no reference system, where they are at rest. The photons are associated
with the massless vector field $V(\phi_{M}\phi^{M})=const.$ The timelike
vector fields can not be associated with the ordinary massive particles
because in the case $\phi_{K}\phi^{K}>0$ there is no frame where $\phi_{0}=0.$
Nevertheless, one can not deny the existence of some substance corresponding
to a timelike field. From the general relativity view point all three kinetic
terms in the Lagrangian (\ref{Lagrangian}) are equally important, as well as
any of the three types of vector fields could describe some sort of matter.

The covariant derivative $\phi_{I;K}$ can be presented as a sum of a symmetric
$G_{IK}$ and an antisymmetric $F_{IK}$\ parts:
\begin{equation}
\phi_{I;K}=G_{IK}+F_{IK},\qquad G_{IK}=\frac{1}{2}\left(  \phi_{I;K}%
+\phi_{K;I}\right)  ,\qquad F_{IK}=\frac{1}{2}\left(  \phi_{I;K}-\phi
_{K;I}\right)  . \label{G_IK, and F_IK}%
\end{equation}
In view of $G_{K}^{L}F_{L}^{K}=0$ the scalar (\ref{Scalar S}) can be presented
in the form
\begin{equation}
S=a\left(  G_{K}^{K}\right)  ^{2}+\left(  b+c\right)  G_{K}^{L}G_{L}%
^{K}+\left(  b-c\right)  F_{K}^{L}F_{L}^{K}.
\label{Scalar S in terms if symm and antisymm}%
\end{equation}
The last term with antisymmetric derivatives is identical to electromagnetism.
It becomes clear in common notations $A_{I}=\phi_{I}/2,$ $\ F_{IK}%
=A_{I;K}-A_{K;I}.$ The bilinear combination of the antisymmetric derivatives
$F_{IK}F^{IK}$ is the same as in electrodynamics. In view of the symmetry of
Christoffel symbols $\Gamma_{IK}^{L}=\Gamma_{KI}^{L},$
\[
A_{I;K}-A_{K;I}=\frac{\partial A_{I}}{\partial x^{K}}-\frac{\partial A_{K}%
}{\partial x^{I}},
\]
and the scalar $F_{IK}F^{IK}$ does not depend on the derivatives of the metric
tensor. On the contrary, the two first terms in
(\ref{Scalar S in terms if symm and antisymm}) with symmetric covariant
derivatives contain not only the components of the metric tensor $g^{IK},$ but
also the derivatives $\frac{\partial g_{IK}}{\partial x^{L}}.$ The difference
between the two terms with symmetric tensors is caused by the curvature of space-time.

\subsection{\label{Regularity in GR}Regularity in general relativity}

In the notations
\begin{equation}
a=A,\qquad b+c=B,\qquad b-c=C \label{a=A, ...}%
\end{equation}
the Lagrangian is
\begin{equation}
L=A\left(  G_{M}^{M}\right)  ^{2}+BG_{MN}G_{\text{ \ \ \ }}^{MN}%
+CF_{MN}F_{\text{ \ \ \ }}^{MN}-V\left(  \phi_{M}\phi^{M}\right)  .
\label{Lagrangian(....)}%
\end{equation}

In plane space-time it is necessary to require $C<0$. Otherwise for a
spacelike vector $\left(  \phi_{0}=0,\text{ }F_{MN}F_{\text{ \ \ \ }}%
^{MN}=\frac{1}{2}\left[  \left(  \text{rot}\overrightarrow{\phi}\right)
^{2}-\frac{1}{c^{2}}\left(  \partial\overrightarrow{\phi}/\partial t\right)
^{2}\right]  \right)  $\ the action could not have a minimum as required by
the least action principle. If $-C\left(  \frac{\partial\overrightarrow{\phi}%
}{\partial t}\right)  ^{2}$ is negative, then it is possible to make the
action negative with an arbitrarily large absolute value via fairly rapid
change of $\overrightarrow{\phi}$ with time (within the considered time
interval), see \cite{Landau-Lifshits}, page 98.

In the regular solutions of the Einstein equations all invariants of the
Riemann curvature tensor are finite. Hence, the invariants of the Ricci tensor
$R_{IK}$ are finite too. By virtue of Einstein equations the requirement of
regularity automatically excludes a possibility to achieve an infinite value
for all the invariants of the energy-momentum tensor $T_{IK}$. In General
Relativity the distribution/motion of matter and the curvature of space-time
are mutually balanced. Practically, there is no need to require $C<0$ in
advance. Necessary restrictions, if any, on the signs of the constants $a,b,$
and $c$ arise as a consequence of the condition of regularity.

The requirement that all the invariants of the Riemann curvature tensor are
finite is a necessary condition of regularity in general relativity.

\subsection{\label{Vec field eq}Vector field equations}

The vector field $\phi_{I}$ obeys the Eiler-Lagrange equations
\begin{equation}
\frac{1}{\sqrt{-g}}\frac{\partial}{\partial x^{L}}\left(  \sqrt{-g}%
\frac{\partial L}{\partial\frac{\partial\phi_{I}}{\partial x^{L}}}\right)
=\frac{\partial L}{\partial\phi_{I}}. \label{Eiler-Lagrange equations}%
\end{equation}
In terms of $a,b,$ and $c$ the variational derivative $\frac{\partial
L}{\partial\frac{\partial\phi_{I}}{\partial x^{L}}}$ is
\[
\frac{\partial L}{\partial\frac{\partial\phi_{I}}{\partial x^{L}}}=2\left(
ag^{IL}\phi_{;K}^{K}+b\phi^{I;L}+c\phi^{L;I}\right)  .
\]
For $\frac{\partial L}{\partial\phi_{I}}$ we have
\[
\frac{\partial L}{\partial\phi_{I}}=\left(  ag^{NK}g^{LM}+bg^{NL}%
g^{KM}+cg^{NM}g^{KL}\right)  \frac{\partial}{\partial\phi_{I}}\phi_{N;K}%
\phi_{L;M}-\frac{\partial}{\partial\phi_{I}}V\left(  g^{NK}\phi_{N}\phi
_{K}\right)  .
\]
$\phi_{I}$ and $\frac{\partial\phi_{I}}{\partial x^{K}}$ are considered as
independent variables in the Lagrangian (\ref{Lagrangian}). In a locally
geodesic reference system (where the Christoffel symbols together with the
derivatives $\frac{\partial g_{IK}}{\partial x^{L}}$ are zeros) $\frac
{\partial}{\partial\phi_{I}}\phi_{N;K}\phi_{L;M}=0,$ and $\frac{\partial
L}{\partial\phi_{I}}=-2V^{\prime}\phi^{I}.$ Here
\begin{equation}
V^{\prime}=\frac{dV}{d\left(  \phi_{L}\phi^{L}\right)  }. \label{V'=}%
\end{equation}
The vector field equations (\ref{Eiler-Lagrange equations}), having a
covariant form
\begin{equation}
a\phi_{;L;I}^{L}+b\phi_{I;L}^{;L}+c\phi_{;I;L}^{L}=-V^{\prime}\phi_{I}
\label{Field eq in terms of a,b, and c}%
\end{equation}
in a locally geodesic system, remain the same in all other reference frames.

In terms of $A,$ $B,$ and $C$
\begin{equation}
AG_{L;I}^{L}+BG_{I;L}^{L}-CF_{\text{ \ }I;L}^{L}=-V^{\prime}\phi_{I}.
\label{Field eq in terms of A,B, and C}%
\end{equation}

There are two independent terms with the symmetric tensor $G$ in
(\ref{Field eq in terms of A,B, and C})\ and one with the antisymmetric tensor
$F.$ The physical origin of the two symmetric terms is connected with the
curvature of space-time. It becomes clear if we set $F_{IK}=0$. Then
$\phi_{I;K;L}=\phi_{K;I;L},$ $G_{I;L}^{L}=g^{KL}\phi_{K;I;L}=\phi_{;I;L}^{L}$
, and (\ref{Field eq in terms of A,B, and C}) reduces to
\begin{equation}
A\phi_{;L;I}^{L}+B\phi_{;I;L}^{L}=-V^{\prime}\phi_{I}.
\label{Field eq in terms of A, and B}%
\end{equation}
The two left terms differ by the order of differentiation. In accordance with
(\ref{Difference between A and B}) the difference between the two terms in
(\ref{Field eq in terms of A, and B}) exists only\ in the curved space-time:
\[
\phi_{;L;I}^{L}-\phi_{;I;L}^{L}=R_{IK}\phi^{K}.
\]
In the flat space-time the Ricci tensor $R_{IK}=0,$ and in case $F_{IK}=0$
there is no physical difference between the two left terms in
(\ref{Field eq in terms of A, and B}).

If the vector field is weak, so that the second and higher derivatives of the
potential $V\left(  \phi_{L}\phi^{L}\right)  $ can be neglected, then the
equations (\ref{Field eq in terms of a,b, and c}) are linear,%
\begin{equation}
a\phi_{;L;I}^{L}+b\phi_{I;L}^{;L}+c\phi_{;I;L}^{L}=-V_{0}^{\prime}\phi
_{I},\qquad V_{0}^{\prime}=V^{\prime}\left(  0\right)  ,
\label{Linear vector field equation}%
\end{equation}
and the principle of superposition takes place, as it should be in the case of
free (non-interacting) fields.

\subsubsection{\label{Proca}Proca equations}

In the particular case $a=0,$ $c=-b$ the field equations
(\ref{Field eq in terms of a,b, and c}) reduce to the Proca equations
\cite{Proca} in the case of no sources:%
\begin{equation}
\left(  F^{IL}\right)  _{;L}=-\frac{V_{0}^{\prime}}{2b}\phi^{I},\qquad
a=0,\text{ \ }c=-b. \label{Proca, a=0, b=-c}%
\end{equation}
Usually $m=\sqrt{-V_{0}^{\text{ }\prime}/b}$ is referred to as the "mass" of a
field. Proca equations are used to describe a free massive spin-1 particle. In
the case of a\ massless field $m=0$ (\ref{Proca, a=0, b=-c}) reduce to the
Maxwell equations.

For any tensor $Q^{IL}$ the scalar $\left(  Q^{IL}\right)  _{;L;I}$ is
symmetric with respect to the lower indexes. Renaming the blind indexes, in
view of the symmetric properties of the Riemann and Ricci tensors%
\begin{align*}
R_{KQTL}  &  =R_{TLKQ}=-R_{LTKQ},\qquad\\
R_{KT}  &  =R_{TK}%
\end{align*}
we have
\begin{equation}
Q_{\text{ \ };L;K}^{LK}-Q_{\text{ \ };K;L}^{LK}=g^{QL}R_{TQLK}Q^{TK}%
+g^{PK}R_{TPLK}Q^{LT}=\left(  R_{KT}-R_{TK}\right)  Q^{TK}=0.
\label{Simmetry against lower indexes}%
\end{equation}
The scalar%
\begin{equation}
\left(  F^{IL}\right)  _{;L;I}=0, \label{F^IL _;I;L=0}%
\end{equation}
because in accordance with (\ref{Simmetry against lower indexes}) $\left(
F^{IL}\right)  _{;L;I}=\left(  F^{IL}\right)  _{;I;L},\ $while for the
antisymmetric tensor $\left(  F^{IL}\right)  _{;L;I}=-\left(  F^{LI}\right)
_{;L;I}=-\left(  F^{IL}\right)  _{;I;L}.$ Thus, it follows from the Proca
equations (\ref{Proca, a=0, b=-c}) that\ in the particular case $a=0,$ $c=-b$
the covariant divergence of the vector $\phi_{I}$ is zero:
\begin{equation}
\phi_{;I}^{I}=0,\qquad a=0,\text{ }c=-b. \label{Lorentz gauge}%
\end{equation}
In electrodynamics $\phi_{;I}^{I}=0$ is referred to as Lorentz gauge. The fact
that $\phi_{;I}^{I}=0$ does not mean that the Proca equations are gauge
invariant. (\ref{Lorentz gauge}) is the consequence of the particular choice
$a=0,$ $c=-b.$

If still\ $c=-b,$ but $a\neq0,$ we have
\begin{equation}
a\left(  \phi_{;L}^{L}\right)  ^{;I}+2b\left(  F^{IL}\right)  _{;L}%
=-V_{0}^{\prime}\phi^{I}. \label{a ne zero c=-b}%
\end{equation}
The Lorentz gauge restriction (\ref{Lorentz gauge}) would automatically
exclude the case $a\neq0$\ from the consideration. Applying $\left(
{}\right)  _{;I}$ to the equation (\ref{a ne zero c=-b}), the term $\sim b$
drops out in view of (\ref{F^IL _;I;L=0}), and, instead of the Lorentz
condition (\ref{Lorentz gauge}), the scalar $\phi_{;I}^{I}$ obeys the
Klein-Gordon equation%
\begin{equation}
\left(  \phi_{;I}^{I}\right)  _{;L}^{;L}\equiv\frac{1}{\sqrt{-g}}%
\frac{\partial}{\partial x^{L}}\left(  \sqrt{-g}g^{KL}\frac{\partial\phi
_{;I}^{I}}{\partial x^{K}}\right)  =-\frac{V_{0}^{\prime}}{a}\phi_{;I}%
^{I},\qquad a\neq0,\text{ }c=-b. \label{eq in case c=-b, a ne zero}%
\end{equation}
In plain space-time \textquotedblleft the additional condition
(\ref{Lorentz gauge}) excludes the part of $\phi^{I}$ belonging to spin
0\textquotedblright\ \cite{Landau-Lifshits4}, page 72. In general relativity
the condition (\ref{Lorentz gauge}) excludes also longitudinal vectors,
existing in curved space-time. As it turns out, these longitudinal vector
fields are just suitable for the description of dark matter and energy.

In the particular case $c=-b$ the field equation
(\ref{eq in case c=-b, a ne zero}) for the divergence $\phi_{;I}^{I}$ does not
contain $b.$ Though the scalar $\phi_{;I}^{I}$ does not depend on $b,$ the
field itself $\phi_{I}$ still obeys the equation (\ref{a ne zero c=-b})
containing $b.$

In the most simple case
\[
a\neq0,\text{ }b=c=0
\]
the field equations (\ref{Linear vector field equation}) reduce to
$\ \frac{\partial\phi_{;L}^{L}}{\partial x^{I}}=0$ $\ $if $V_{0}^{\prime}=0$
(massless field)$,$ and to
\begin{equation}
\phi_{I}=-\frac{a}{V_{0}^{\prime}}\frac{\partial\phi_{;L}^{L}}{\partial x^{I}%
}\text{ },\text{ \ \ \ }a\neq0,\text{ }b=c=0, \label{fi_I = ...   b=c=0}%
\end{equation}
if $V_{0}^{\prime}\neq0$ (massive field)$.$

The covariant divergence $\phi_{;L}^{L}$ of a zero-mass field remains constant
through the whole space-time:
\begin{equation}
\phi_{;L}^{L}\equiv\phi_{0}^{\prime}=const,\text{ \ \ }a\neq0,\text{
}b=c=0,\text{ \ \ \ }V^{\prime}=0. \label{div fi =fi'_0}%
\end{equation}
The fact that \textquotedblleft the gauge-fixing term exactly behaves as a
cosmological constant throughout the history of the universe, irrespective of
the background evolution\textquotedblright\ had been mentioned by Jimenez and
Maroto \cite{Jimenez2}. The massive field $\phi_{I}$ has a potential: it is a
gradient of the scalar $\phi_{;L}^{L}$ obeying the equation
(\ref{eq in case c=-b, a ne zero})$.$

\subsubsection{Einstein-aether models}

Vector field $\phi_{K}$ considered as the gradient of a scalar field $\Phi$,
$\left(  \phi_{K}=\partial\Phi/\partial x^{K}\right)  ,$ was used in a scalar
variant of the Einstein-aether theory, see a recent paper by Zahra Haghani,
Tiberiu Harko, et al \cite{Haghani}, providing a brief comprehensive review of
the topic. Einsten-aether theories \cite{Jacobson},\cite{Jacobson1} consider
phase transitions with spontaneous violation of Lorentz symmetry by a vector
field whose non-zero vacuum expectation value plays the role of the order
parameter. "Einstein-aether" is a kind of bumblebee models oriented mainly on
timelike vector fields. It is not clear how to associate a timelike vector
field with a massive quantum particle of ordinary matter, because there is no
reference frame where such a particle could be at rest. The word "aether" in a
title reflects the situation that a timelike vector field should correspond to
something different from the ordinary matter. \ 

A tipical action in the Einsten-aether theory \cite{Carroll}%
-\cite{Armendariz-Picon2} is a particular case of $\left(
\ref{Most general Lagrangian}\right)  $ with $J^{IKLM}=0,$ and
\begin{equation}
K^{IKLM}=c_{0}g^{IK}g^{LM}+c_{1}g^{IM}g^{KL}+c_{2}g^{IL}g^{KM}+c_{3}\phi
^{I}\phi^{K}g^{LM}. \label{EA K^IKLM}%
\end{equation}
In \cite{Haghani} the potential $V\left(  \phi_{I}\phi^{I}\right)  $ is taken
as $V=\lambda\left(  \phi_{I}\phi^{I}\pm1\right)  ,$ where $\lambda$ is a
Lagrange-Eiler multiplier. The sign being chosen to enforce the vector field
$\phi_{I}$ to be timelike. Another commonly used example for the potential is
a smooth quadratic function $V=\frac{1}{2}\lambda\left(  \phi_{I}\phi^{I}%
\pm1\right)  ^{2}.$

If the symmetry breaking vector $\phi_{I}$ is chosen as a gradient of a
scalar, $\phi_{I}=\frac{\partial\Phi}{\partial x^{I}}$, then it is possible to
consider the potential $V$ as a function of the scalar $\Phi$ only,
$V=V\left(  \Phi\right)  .$ In this case a Lagrangian contains not only $\Phi$
and $\frac{\partial\Phi}{\partial x^{I}},$ but also second derivatives of
$\Phi$. From my point of view, a vector field approach is more convenient than
the scalar one. The equations become more simple, while their solutions are
more general. Though in the particular case $a\neq0,$ $b=c=0$ the vector
$\phi_{I}$ is a gradient of a scalar, see $\left(  \ref{fi_I = ... b=c=0}%
\right)  ,$ this scalar is a covariant divergence $\phi_{;L}^{L}.$ The
Lagrangian contains only $\phi_{I}$ and $\phi_{;K}^{I}$.

In the state of broken symmetry there are 4 constants $\left(  c_{0}%
,c_{1},c_{2},c_{3}\right)  $ in $\left(  \ref{EA K^IKLM}\right)  $ instead of
3 constants $\left(  a,b,c\right)  $ in (\ref{Lagrangian})$.$ If a vector
field is small compared to its vacuum expectation value, and $V_{0}^{\prime
}\equiv\frac{dV\left(  \phi^{K}\phi_{K}\right)  }{d\left(  \phi^{K}\phi
_{K}\right)  }\left\vert _{\phi^{K}\phi_{K}=0}\right.  \neq0,$ then $V\left(
\phi^{K}\phi_{K}\right)  =V\left(  0\right)  +V_{0}^{\prime}\phi^{K}\phi_{K},$
and the second and higher derivatives of $V\left(  \phi^{K}\phi_{K}\right)  $
can be omitted. While the field is small, the symmetry remains unbroken, and
there is no need in the fourth term in $\left(  \ref{EA K^IKLM}\right)  .$ On
the contrary, the full scale bumblebee models, including the Einsten-aether
theory, provide a solid basis for analyzing phase transitions with spontaneous
symmetry breaking in a strongly compressed state.

\subsubsection{Nongauge longitudinal vector field in plane geometry}

\begin{figure}
[ptbh]
\begin{center}
\includegraphics[
trim=0.000000in 0.000000in 0.005288in 0.001508in,
height=6.021cm,
width=9.0062cm
]%
{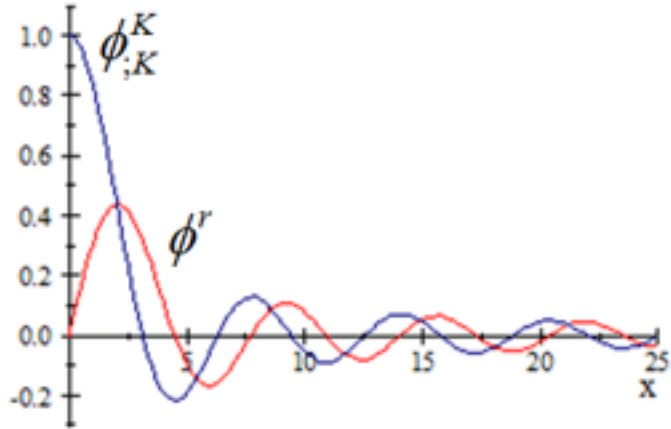}%
\caption{ $\phi^{r}\sim\frac{1}{x}\left(  \frac{1}{x}\sin x-\cos x\right)  $
and $\phi_{;M}^{M}\sim\frac{1}{x}\sin x$ }%
\label{fig2}%
\end{center}
\end{figure}

In the plain centrally symmetric metric%
\[
ds^{2}=dx^{0\text{ }2}-dr^{2}-r^{2}\left(  d\theta^{2}+\sin^{2}\theta
d\varphi^{2}\right)  ,\qquad g=-r^{4}\sin^{2}\theta
\]
$g^{rr}=-1,$ and the equation (\ref{eq in case c=-b, a ne zero})\ for a static
longitudinal vector field, depending only on the distance $r$ from the center,
is%
\begin{equation}
\frac{1}{r^{2}}\frac{d}{dr}r^{2}\frac{d\phi_{;L}^{L}}{dr}=\frac{V_{0}^{\prime
}}{a}\phi_{;L}^{L}. \label{eq for div fi spherical}%
\end{equation}
The center $r=0$ is a singular point of the spherical coordinate system. Like
a pole on the globe, the singularity of a coordinate system has nothing to do
with physical properties of matter. The finite, regular at $r\rightarrow
0,$\ solution$\allowbreak$ of the equation (\ref{eq for div fi spherical})%

\begin{equation}
\phi_{;K}^{K}=\phi_{0}^{\prime}\frac{\sin\left(  mr\right)  }{mr},\qquad
m=\sqrt{-\frac{V_{0}^{\prime}}{a}} \label{fi^K_;K=}%
\end{equation}
(blue curve in Figure \ref{fig2}) exists if
\begin{equation}
\frac{V_{0}^{\prime}}{a}<0. \label{V'/(A+B)<0}%
\end{equation}
\ \ Restriction (\ref{V'/(A+B)<0}) is a necessary condition of regularity for
a spacelike vector field.\ 

According to (\ref{fi_I = ... b=c=0}) the nonzero component of the vector
$\phi^{I}$ is $\phi^{r}$ :
\begin{equation}
\phi^{r}=\frac{\phi_{0}^{\prime}}{mr}\left[  \frac{1}{mr}\sin\left(
mr\right)  -\cos\left(  mr\right)  \right]  . \label{fi^r=}%
\end{equation}
Radial dependences of $\phi_{;K}^{K}$\ and $\phi^{r}$ are shown in Figure
\ref{fig2}. The field is spacelike and longitudinal: it is directed along and
depends upon the same coordinate $r$.

Boundary conditions for the equation (\ref{eq for div fi spherical}),
separating the regular solution at $r\rightarrow0,$ are%
\begin{equation}
\phi_{;L}^{L}\left(  0\right)  \equiv\phi_{0}^{\prime},\qquad\frac{d\phi^{r}%
}{dr}\left\vert _{r=0}\right.  =\frac{1}{3}\phi_{0}^{\prime}%
.\label{boud cond for div fi}%
\end{equation}
(\ref{eq for div fi spherical}) is a linear uniform equation, and the constant
of integration $\phi_{0}^{\prime}$ remains arbitrary.

In plain geometry $g^{00}=1,$ and in the case of a longitudinal timelike
vector field the Klein-Gordon equation (\ref{eq in case c=-b, a ne zero}) is
simply
\begin{equation}
\frac{d^{2}\phi_{;L}^{L}}{\left(  dx^{0}\right)  ^{2}}=-\frac{V_{0}^{\prime}%
}{a}\phi_{;L}^{L}. \label{eq for div fi time-like}%
\end{equation}
Its solution $\phi_{;L}^{L}=\phi_{0}^{\prime}\sin\left(  \sqrt{\frac
{V_{0}^{\prime}}{a}}\left(  x^{0}-x_{0}^{0}\right)  \right)  $ is finite if
\begin{equation}
\frac{V_{0}^{\prime}}{a}>0. \label{V'/a >0}%
\end{equation}
$\phi_{0}^{\prime}$ and $x_{0}^{0}$ are constants of integration.

Inequality (\ref{V'/a >0}) is a necessary condition of regularity for a
timelike massive vector field. It is just the opposite to the one for a
spacelike massive field (\ref{V'/(A+B)<0}).

In plain space-time conditions of regularity determine the sign of the ratio
$\frac{V_{0}^{\prime}}{a},$ which appears different for spacelike and timelike
fields. In plain space-time the specific sign of $a$ is not restricted by the
requirement of regularity. However, there is no reason why the sign of $a$
should be different for spacelike and timelike vector fields. Actually it
turns out that the sign of $a,$%
\begin{equation}
a<0, \label{a <0}%
\end{equation}
is restricted by the requirement of regularity in a curved space-time, see
section \ref{Massless+massive} below. Then, as it follows from
(\ref{V'/(A+B)<0}, \ref{V'/a >0}, and \ref{a <0}), the regular solutions
exist, if $V_{0}^{\prime}$ is positive for a spacelike field, and negative --
for a timelike one.

\subsection{\label{En-mom tensor}Energy-momentum tensor}

Using the identity%

\[
\delta g_{IK}=-g_{KM}g_{IN}\delta g^{NM},
\]
the energy-momentum tensor can be expressed as:
\begin{equation}
T_{IK}=\frac{2}{\sqrt{-g}}\left[  \frac{\partial\sqrt{-g}L}{\partial g^{IK}%
}+g_{MI}g_{NK}\frac{\partial}{\partial x^{L}}\left(  \sqrt{-g}\frac{\partial
L}{\partial\frac{\partial g_{MN}}{\partial x^{L}}}\right)  \right]  .
\label{Tik general}%
\end{equation}
It differs from (94.4) in \cite{Landau-Lifshits}, where the Lagrangian is a
function of $g^{IK}$ and $\frac{\partial g^{IK}}{\partial x^{L}}$. The form
(\ref{Tik general}) is more convenient when $g^{IK}$ and $\frac{\partial
g_{IK}}{\partial x^{L}}$ are considered as independent variables in the
Lagrangian (\ref{Lagrangian}). In view of%
\[
\frac{2}{\sqrt{-g}}\frac{\partial\sqrt{-g}}{\partial g^{IK}}=-g_{IK}.
\]
in a locally geodesic system (where $\frac{\partial g^{IK}}{\partial x^{L}}%
=0$) the energy-momentum tensor can be written as follows:
\begin{equation}
T_{IK}=-g_{IK}L+2\frac{\partial L}{\partial g^{IK}}+2g_{MI}g_{NK}\left(
\frac{\partial L}{\partial\frac{\partial g_{MN}}{\partial x^{L}}}\right)
_{;L}. \label{Tik covariant}%
\end{equation}
It is worth mentioning that $\frac{\partial g^{IK}}{\partial x^{L}}$ should be
set to zero\ after the variational differentiations $\frac{\partial
L}{\partial g^{IK}}$\ and $\frac{\partial L}{\partial\frac{\partial g_{MN}%
}{\partial x^{L}}}$ are done. In terms of symmetric and antisymmetric tensors
(\ref{G_IK, and F_IK}) we find:
\begin{equation}
\frac{\partial L}{\partial g^{IK}}=2\left(  AG_{L}^{L}G_{IK}+BG_{K}^{L}%
G_{IL}+CF_{I}^{\text{ \ \ }L}F_{KL}\right)  -V^{\prime}\phi_{I}\phi_{K},
\label{dL/dg=}%
\end{equation}%
\begin{equation}
\frac{\partial L}{\partial\frac{\partial g_{MN}}{\partial x^{L}}}=-A\phi
_{;P}^{P}\left(  g^{LN}\phi^{M}+g^{LM}\phi^{N}-g^{NM}\phi^{L}\right)
-B\left(  G^{LN}\phi^{M}+G^{LM}\phi^{N}-G^{MN}\phi^{L}\right)  .
\label{dL/d(dg/dx)=}%
\end{equation}
The tensor (\ref{dL/d(dg/dx)=}) is presented in a symmetric form against the
indexes $M,N.$

Substituting (\ref{dL/dg=}) and (\ref{dL/d(dg/dx)=}) into (\ref{Tik covariant}%
), we find the following covariant expression for the energy-momentum tensor:
\begin{equation}%
\begin{array}
[c]{l}%
T_{IK}=-g_{IK}L+2V^{\prime}\phi_{K}\phi_{I}+2Ag_{IK}\left(  G_{M}^{M}\phi
^{L}\right)  _{;L}+2B\left[  \left(  G_{IK}\phi^{L}\right)  _{;L}-G_{K}%
^{L}F_{IL}-G_{I}^{L}F_{KL}\right] \\
+2C\left(  2F_{\text{ }I}^{L}F_{LK}-F_{\text{ \ }K;L}^{L}\phi_{I}-F_{\text{
}I;L}^{L}\phi_{K}\right)  .
\end{array}
\label{General Tik Covari}%
\end{equation}

The vector field equations (\ref{Field eq in terms of A,B, and C})\ were used
to reduce $T_{IK}$ to a rather simple form (\ref{General Tik Covari}).

\subsubsection{\label{Checking}Checking the zero of the covariant divergence
$\ T_{I;K}^{K}=0$}

The correctness of the energy-momentum tensor (\ref{General Tik Covari}) is
confirmed by demonstration that the covariant divergence

$T_{I;K}^{K}=-L_{;I}+2\left(  V^{\prime}\phi^{K}\phi_{I}\right)
_{;K}+2A\left(  G_{M}^{M}\phi^{L}\right)  _{;L;I}+2B\left[  \left(  G_{IK}%
\phi^{L}\right)  _{;L}-G^{LK}F_{IL}-G_{I}^{L}F_{\text{ \ \ }L}^{K}\right]
_{;K}+2C\left(  2F_{\text{ }I}^{L}F_{L}^{\text{ \ }K}-F_{\text{ \ \ \ }%
;L}^{LK}\phi_{I}-F_{\text{ }I;L}^{L}\phi^{K}\right)  _{;K}$ \newline is zero
\cite{Meier2}.

Using the vector field equations (\ref{Field eq in terms of A,B, and C}),
$T_{I;K}^{K}$ can be presented as
\begin{equation}
T_{I;K}^{K}=Aa_{I}+Bb_{I}+Cc_{I}. \label{Aai+Bbi+Cci}%
\end{equation}
The coefficients $A,B,$ and $C$ are arbitrary constants. However, it doesn't
mean that the three vectors $a_{I},$ $b_{I},$ and $c_{I}$ in
(\ref{Aai+Bbi+Cci}) are zeros separately. These vectors are reduced to a
similar form (see \cite{Meier2} for details):%

\begin{equation}
a_{I}=2\left(  \phi_{;K;I}^{K}\phi_{L}-\phi_{;K;L}^{K}\phi_{I}\right)  ^{;L}.
\label{ai=}%
\end{equation}
\begin{equation}
b_{I}=2\left(  G_{I;K}^{K}\phi_{L}-G_{L;K}^{K}\phi_{I}\right)  ^{;L},
\label{bi=}%
\end{equation}%
\begin{equation}
c_{I}=2\left(  F_{\text{ \ }L;K}^{K}\phi_{I}-F_{\text{ \ }I;K}^{K}\phi
_{L}\right)  ^{;L}. \label{ci=}%
\end{equation}
The covariant divergence of the energy-momentum tensor (\ref{Aai+Bbi+Cci})
with $a_{I},b_{I},$ and $c_{I},$ given by (\ref{ai=}-\ref{ci=}), is evidently
zero due to the vector field equations (\ref{Field eq in terms of A,B, and C}).

The covariant field equations (\ref{Field eq in terms of a,b, and c}) and the
energy-momentum tensor (\ref{General Tik Covari}) describe the behavior of
vector fields in the background of any arbitrary given metric $g_{IK}$
\cite{Meier2}$.$ If the back reaction of the field on the curvature of
space-time is essential, then the metric obeys the Einstein equations
\begin{equation}
R_{IK}-\frac{1}{2}g_{IK}R+\Lambda g_{IK}=\varkappa T_{IK}
\label{Einstein equations}%
\end{equation}
with (\ref{General Tik Covari}) added to $T_{IK}.$ Here $\Lambda$ and
$\varkappa$ are the cosmological and gravitational constants, respectively.
With account of back reaction the field equations
(\ref{Field eq in terms of a,b, and c}) are not independent. They follow from
the Einstein equations (\ref{Einstein equations}) with $T_{IK}$
(\ref{General Tik Covari}) due to the Bianchi identities.

\subsubsection{\label{Ordinary matter and the dark sector}Ordinary matter and
dark sector}

The ordinary matter enters the Einstein equations via the well known
energy-momentum tensor of macroscopic objects.%
\begin{equation}
T_{\text{om }IK}=\left(  \varepsilon+p\right)  u_{I}u_{K}-pg_{IK}
\label{Tom_IK}%
\end{equation}
The energy $\varepsilon,$ pressure $p,$ (and, generally speaking, temperature
$T$) of the ordinary matter obey the equation of state. So far there is no
evidence of any direct interaction between dark and ordinary matter other than
via gravitation. The gravitational interaction is described by Einstein
equations (\ref{Einstein equations}) with
\begin{equation}
T_{IK}=T_{\text{dark }IK}+T_{\text{om }IK}. \label{T_IK=dm+om}%
\end{equation}
The general expression (\ref{General Tik Covari}) for the energy-momentum
tensor of a vector field describes (but is not limited to) the vector
particles, which are also the ordinary matter. In order to describe the dark
sector via vector fields it is reasonable to separate from $T_{IK}$
(\ref{General Tik Covari}) the part $T_{\text{dark }IK}$ which does not relate
to the ordinary matter.

The Proca equations (\ref{Proca, a=0, b=-c}) are associated with a spin-1
particle -- a quantum of the ordinary matter. The particular choice of the
parameters $b=-c$ ensures the Lorentz condition (\ref{Lorentz gauge}), which
allows to avoid \textquotedblleft the difficulties of negative contribution to
the energy\textquotedblright\ \cite{Bogolubov Shirkov}. However, in curved
space-time the energy is not a scalar, and its sign is not invariant against
the arbitrary coordinate transformations. As a result of the Lorentz gauge
restriction the terms $\sim a$ drop out from the Lagrangian (\ref{Lagrangian}%
), from the vector field equations (\ref{Field eq in terms of a,b, and c}),
and from the energy-momentum tensor (\ref{General Tik Covari}).

To avoid the double contribution of particles of the ordinary matter, it is
reasonable to consider $a\neq0,$ and set $b=c=0$. As simple a Lagrangian as
possible%
\begin{equation}
L_{\text{dark}}=a\left(  \phi_{;M}^{M}\right)  ^{2}-V\left(  \phi^{L}\phi
_{L}\right)  \label{L dark}%
\end{equation}
turns out to be an adequate tool for macroscopic description of the dark
sector. Accordingly, the equations (\ref{Field eq in terms of a,b, and c}) and
the energy-momentum tensor (\ref{General Tik Covari}) reduce to%
\begin{align}
a\frac{\partial\phi_{;M}^{M}}{\partial x^{I}}  &  =-V^{\prime}\phi
_{I},\label{Field eqs b=c=0}\\
T_{\text{dark }IK}  &  =g_{IK}\left[  a(\phi_{;M}^{M})^{2}+V\right]
+2V^{\prime}\left(  \phi_{I}\phi_{K}-g_{IK}\phi^{M}\phi_{M}\right)  .
\label{Tik b=c=0}%
\end{align}

Gravity is currently considered the basic interaction of cosmic objects in the
scale of galaxies and larger. The three types of vector fields (massless,
massive spacelike, and massive timelike) can be of different physical nature.

The energy-momentum tensor (\ref{Tik b=c=0}) of a zero-mass $\left(
V^{\prime}=0\right)  $ vector field reduces to%
\begin{equation}
T_{\text{(0) }IK}=g_{IK}\left(  a\phi_{0}^{\prime\text{ }2}+V_{\text{(0)}%
}\right)  . \label{T_(0) IK= massless field}%
\end{equation}
$V_{\text{(0)}}$ is the constant value of the potential of the massless field.
$T_{\text{(0) }IK}$ acts in the Einstein equations (\ref{Einstein equations})
as a simple addition to the cosmological constant, changing $\Lambda$ to
\begin{equation}
\widetilde{\Lambda}=\Lambda-\varkappa\left(  a\phi_{0}^{\prime\text{ }2}%
+V_{0}\right)  . \label{Lambda with tilda}%
\end{equation}
$\phi_{0}^{\prime}$ is the constant divergence of the zero-mass vector field
(\ref{div fi =fi'_0}).

It is convenient to consider the values $V\left(  0\right)  $ of the
potentials of massive fields as already included into $\widetilde{\Lambda}:$
\[
V_{0}=V_{\text{(0)}}+V_{\text{(s)}}\left(  0\right)  +V_{\text{(t)}}\left(
0\right)  ,
\]
so that the power series of the potentials $V\left(  \phi^{M}\phi_{M}\right)
$ of massive fields start with $V_{0}^{\prime}\phi^{M}\phi_{M}:$
\[
V\left(  \phi^{M}\phi_{M}\right)  =V_{0}^{\prime}\phi^{M}\phi_{M}+O\left(
\left(  \phi^{M}\phi_{M}\right)  ^{2}\right)  .
\]
In the case of weak vector fields the second and higher derivatives of the
potentials $V\left(  \phi_{L}\phi^{L}\right)  $ can be neglected, and the
energy-momentum tensor of a massive field is%
\begin{equation}
T_{\text{dark }I}^{K}=a(\phi_{;M}^{M})^{2}\delta_{I}^{K}+V_{0}^{\prime}\left(
2\phi_{I}\phi^{K}-\delta_{I}^{K}\phi^{M}\phi_{M}\right)  ,\qquad V_{0}%
^{\prime}\equiv\frac{dV\left(  \phi^{M}\phi_{M}\right)  }{d\left(  \phi
^{M}\phi_{M}\right)  }\left\vert _{_{\phi^{M}\phi_{M}=0}}\right.  .
\label{T_dark weak field}%
\end{equation}

In general, it could be necessary to consider both independent vectors --
$\phi_{\text{(s)}}^{M}$ for a spacelike$,$ and $\phi_{\text{(t)}}^{M}$ for a
timelike -- massive fields with different potentials $V_{\left(
\text{s}\right)  }\left(  \phi_{\text{(s)}}^{M}\phi_{\text{(s) }M}\right)  $
and $V_{\text{(t)}}\left(  \phi_{\text{(t)}}^{M}\phi_{\text{(t) }M}\right)  $.
As far as the energy-momentum tensor of a massless field (dark energy) is
included into $\widetilde{\Lambda}$ (\ref{Lambda with tilda})$,$ the remaining
energy-momentum tensor $T_{\text{dark }IK}$ of massive fields is the sum of
two tensors
\begin{equation}
T_{\text{dark }IK}=T_{\text{(s) }IK}+T_{\text{(t)}IK}, \label{T_IK dark sum}%
\end{equation}
corresponding to $\phi_{\text{(s)}}^{M}$ and $\phi_{\text{(t)}}^{M},$ \ respectively.

In the scale of a galaxy ($\sim10$ kpc) the spacelike vector field
$(\phi_{\text{(s)}}^{L}\phi_{\text{(s) }L}<0)$ dominates. It is responsible
for the plateau in galaxy rotation curves, see section \ref{Rotation curves}.
The timelike field $(\phi_{\text{(t)}}^{L}\phi_{\text{(t) }L}>0)$ dominates at
the scales much larger than the distance between the galaxies, where the
universe can be considered as uniform and isotropic. The timelike field
displays the repulsive elasticity. Together with the zero-mass vector field
(dark energy) and the ordinary matter it gives rise to a variety of possible
regular scenarios of evolution of the Universe and rules out the problem of
fine tuning, see section \ref{Cosmology}. In particular, the singular Big Bang
turns into a regular inflation-like bounce with accelerated expansion at late times.

It would be interesting to trace how the additional attraction of the
spacelike dark matter, dominating in the galaxy scale, transforms into the
elastic repulsion of the timelike dark matter, dominating in the scale of the
Universe. Both types of massive fields $\phi_{\text{(s)}}^{M}$ and
$\phi_{\text{(t)}}^{M}$ are supposed to be active in the intermediate region,
so the energy-momentum tensor of the dark sector should be the sum
(\ref{T_IK dark sum}).

The study of the structure of the Universe in the intermediate range (Mpc to
hundred Mpcs) had been initiated in the pioneering papers by Zel'dovich
\cite{Zel'dovich}. Continuous research by his followers shows that dark energy
and dark matter significantly affect the structural dynamics of galaxies and
clusters in this range, see a review by Gurbatov, Saichev, and Shandarin
\cite{Shandarin}. Utilizing the energy-momentum tensor (\ref{T_IK dark sum},
\ref{Tik b=c=0}) in the analysis of the large scale structure of the Universe
would allow to avoid unnecessary model assumptions.

\section{\label{Rotation curves}Galaxy rotation curves}

The description of the dark sector via vector fields allows to derive the
galaxy rotation curves directly from the first principles within the minimal
Einstein's general relativity \cite{Meierovich}.

\begin{figure}
[ptbh]
\begin{center}
\includegraphics[
height=7.9828cm,
width=13.7257cm
]%
{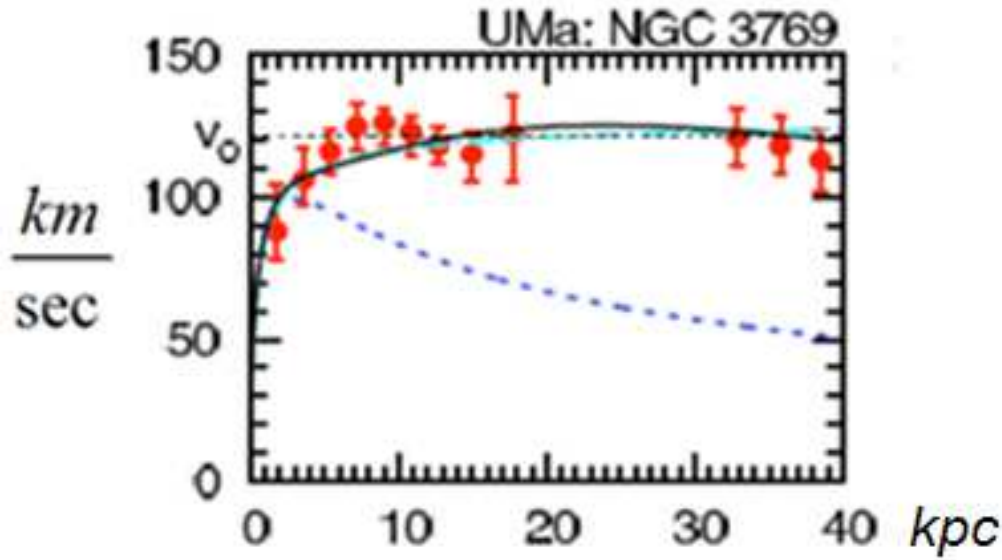}%
\caption{ Rotation curve $V\left(  r\right)  $ of a spiral galaxy in the Ursa
Major cluster (UMa). In the New General Catalogue of Nebulae and Clusters of
Stars (NGC)\cite{Dreyer} its number is 3769. The vertical axis is the velocity
$V$ in $km/\sec$, and the horizontal axis is the distance $r$\ from the center
of the galaxy in $kpc$. Dots with error bars are observations. Solid curve is
fitting by empirical MOND (modified Newton's dynamics) \cite{Milgrom}%
,\cite{Brownstein}. Dashed line is the Newton's $V(r)\sim1/\sqrt{r}.$ }%
\label{fig3}%
\end{center}
\end{figure}

The velocity $V$ of a star, orbiting around the center of a galaxy and
satisfying the balance between the centrifugal $\frac{V^{2}}{r}$ and
centripetal $\frac{GM\left(  r\right)  }{r^{2}}$ accelerations, should
decrease with the radius $r$ of its orbit as $V\left(  r\right)  \sim
1/\sqrt{r}$ at $r\rightarrow\infty.$ However, the numerous observed
dependences $V\left(  r\right)  ,$ named galaxy rotation curves, practically
remain constant at far periphery of a galaxy. An example is presented in
Figure \ref{fig3}.
It was a fundamental problem, because the general relativity reduces to the
Newton's theory in the limit of non-relativistic velocities and weak gravitation.

\subsection{\label{Brief history}Benchmarks in history}

The \textquotedblleft galaxy rotation curves\textquotedblright\ problem
appeared after J. H. Oort discovered the galactic halo, a group of stars
orbiting the Milky Way outside the main disk \cite{Oort 1924}. In 1933, F.
Zwicky \cite{Zwicky 1933} postulated "missing mass" to account for the orbital
velocities of galaxies in clusters. Persistent investigations by V. Rubin and
colleagues \cite{V Rubin},\cite{V Rubin2} in seventies practically dispelled
the skepticism about the existence of dark matter on the periphery of the galaxies.

Among numerous attempts to solve the problem of galaxy rotation curves the
most discussed one is the empirical explanation named MOND (Modified Newtonian
Dynamics), proposed by Milgrom back in 1983 \cite{Milgrom},\cite{Milgrom2}.
For a relativistic justification of MOND Bekenstein \cite{Bekenstein}, Sanders
\cite{Sanders}, Brownstein and Moffat \cite{Brownstein},\cite{Moffat}
introduce additional scalar, vector, or tensor fields. Though these (and many
others) empirical improvements of MOND are able to fit a large number of
samples for about a hundred galaxies, the concern still remains. So far we had
neither self-consistent description of the dark sector as a whole, nor direct
derivation of MOND from the first principles within the Einstein's general
relativity. The survey \cite{Fameay and McGaugh} by Benoit Famaey and Stacy
McGaugh and\ a recent review article \cite{BERNABEI et al} by R. Bernabei, P.
Belli et al reflect the current state of research and contain the
comprehensive lists of references.

\subsection{\label{Derivation}Rotation curve driven by a massive vector field}

Applying general relativity to the galaxy rotation problem it is reasonable to
consider a static centrally symmetric metric%

\begin{equation}
ds^{2}=g_{IK}dx^{I}dx^{K}=e^{\nu\left(  r\right)  }\left(  dx^{0}\right)
^{2}-e^{\lambda\left(  r\right)  }dr^{2}-r^{2}d\Omega^{2}
\label{Centr symm metric}%
\end{equation}
with two functions $\nu\left(  r\right)  $ and $\lambda\left(  r\right)  $
depending on only one coordinate - circular radius $r$. Real distribution of
stars and planets in a galaxy is neither static, nor centrally symmetric.
However this simplification facilitates analyzing the problem and allows to
display the main results analytically. If a galaxy is concentrated around a
supermassive black hole, the deviation from the central symmetry caused by the
peripheral stars is small.

In the background of the centrally symmetric metric (\ref{Centr symm metric})
the vector $\phi^{I}$ is longitudinal. In accordance with the field equation
(\ref{Field eqs b=c=0}) its only non-zero component $\phi^{r}$ depends on $r.$
The covariant divergence is%
\begin{equation}
\phi_{;M}^{M}=\frac{1}{\sqrt{-g}}\frac{\partial\left(  \sqrt{-g}\phi
^{M}\right)  }{\partial x^{M}}=\frac{d\phi^{r}}{dr}+\left(  \frac{2}{r}%
+\frac{\lambda^{\prime}+\nu^{\prime}}{2}\right)  \phi^{r}. \label{div(fi)}%
\end{equation}

In the \textquotedblleft dust matter\textquotedblright\ approximation $p=0,$
and the only nonzero component of the energy-momentum tensor (\ref{Tom_IK}) is
$T_{\text{om }00}=\varepsilon g_{00}$. Whatever the distribution of the
ordinary matter $\varepsilon\left(  r\right)  $ is, the covariant divergence
$T_{\text{om }I;K}^{K}$ is automatically zero. In the dust
matter\ approximation the curving of space-time by ordinary matter is taken
into account, but the back reaction of the gravitational field on the
distribution of matter is ignored. If $p=0$ the energy $\varepsilon\left(
r\right)  $ is considered as a given function.

In the space-time with metric (\ref{Centr symm metric}) the energy-momentum
tensor (\ref{T_dark weak field}) of a\ weak spacelike longitudinal vector
field is
\[
T_{\text{dark }I}^{K}=\delta_{I}^{K}\left\{
\begin{array}
[c]{c}%
a(\phi_{;M}^{M})^{2}-V_{0}^{\prime}e^{\lambda}\phi^{r2},\qquad I=r,\\
a(\phi_{;M}^{M})^{2}+V_{0}^{\prime}e^{\lambda}\phi^{r2},\qquad I\neq r.
\end{array}
\right.
\]
In the scale of galaxies the role of expansion of the Universe as a whole is
negligible, and one can omit $\widetilde{\Lambda}$ (\ref{Lambda with tilda})
in the Einstein equations. In the dust matter approximation the Einstein
equation are:%
\begin{align}
-e^{-\lambda}\left(  \frac{1}{r^{2}}-\frac{\lambda^{\prime}}{r}\right)
+\frac{1}{r^{2}}  &  =\varkappa T_{0}^{0}=\varkappa\left[  a(\phi_{;M}%
^{M})^{2}+V_{0}^{\prime}e^{\lambda}\left(  \phi^{r}\right)  ^{2}%
+\varepsilon\right] \label{0-0 eq}\\
-e^{-\lambda}\left(  \frac{\nu^{\prime}}{r}+\frac{1}{r^{2}}\right)  +\frac
{1}{r^{2}}  &  =\varkappa T_{r}^{r}=\varkappa\left[  a(\phi_{;M}^{M}%
)^{2}-V_{0}^{\prime}e^{\lambda}\left(  \phi^{r}\right)  ^{2}\right]
\label{r-r eq}\\
-\frac{1}{2}e^{-\lambda}\left(  \nu^{\prime\prime}+\frac{\nu^{\prime2}}%
{2}+\frac{\nu^{\prime}-\lambda^{\prime}}{r}-\frac{\nu^{\prime}\lambda^{\prime
}}{2}\right)   &  =\varkappa\left[  a(\phi_{;M}^{M})^{2}+V_{0}^{\prime
}e^{\lambda}\left(  \phi^{r}\right)  ^{2}\right]  ,\qquad I,K\neq0,r.
\label{other eqs}%
\end{align}
See \cite{Landau-Lifshits}, page 382 for the derivation of the left-hand
sides. The prime $^{\prime}$ stands for $\frac{d}{dr},$ except $V^{\prime
}=\frac{dV\left(  \phi_{M}\phi^{M}\right)  }{d\left(  \phi_{M}\phi^{M}\right)
}.$ Among the four equations (\ref{Field eqs b=c=0}), (\ref{0-0 eq}%
-\ref{other eqs}) for the unknowns $\phi^{r},\lambda,$ and $\nu$ any three are independent.

Extracting (\ref{r-r eq}) from (\ref{0-0 eq}) we get a relation%
\begin{equation}
\nu^{\prime}+\lambda^{\prime}=\varkappa re^{\lambda}\left[  2e^{\lambda
}\left(  \phi^{r}\right)  ^{2}V_{0}^{\prime}+\varepsilon\right]  .
\label{Equation for nu'+lambda'}%
\end{equation}
With account of (\ref{div(fi)}) and (\ref{Equation for nu'+lambda'}) the
vector field equation (\ref{Field eqs b=c=0}) takes the form%
\begin{equation}
\left[  \left(  \phi^{r}\right)  ^{\prime}+\left(  \frac{2}{r}+\varkappa
re^{2\lambda}\left(  \phi^{r}\right)  ^{2}V_{0}^{\prime}+\frac{1}{2}\varkappa
re^{\lambda}\varepsilon\right)  \phi^{r}\right]  ^{\prime}=\frac{V^{\prime}%
}{a}e^{\lambda}\phi^{r} \label{Field eq spherical symm}%
\end{equation}
The equations (\ref{Equation for nu'+lambda'}) and
(\ref{Field eq spherical symm}) are derived with no assumptions concerning the
strength of the gravitational field.

Omitting the second and higher derivatives of the potential $V(\phi_{M}%
\phi^{M}),$ in the dust matter approximation $\left(  p=0\right)  $ we get
from (\ref{0-0 eq}) the following expression for $\nu^{\prime}:$
\begin{equation}
\nu^{\prime}=-a\varkappa re^{\lambda}\left[  m^{2}e^{\lambda}\left(  \phi
^{r}\right)  ^{2}+(\phi_{;M}^{M})^{2}\right]  +\frac{e^{\lambda}-1}{r}.
\label{nu'=-lambda' +...}%
\end{equation}
\ Here $m^{2}=-\frac{V_{0}^{\prime}}{a}$ is the squared mass of the vector
field. The sign minus is in conjunction with the necessary condition of
regularity (\ref{V'/(A+B)<0}) for a spacelike vector field.

In a static centrally symmetric gravitational field$\ \nu^{\prime}$ determines
the centripetal acceleration of a particle (\cite{Landau-Lifshits}, page 323).
Without dark matter $\phi^{r}=0$ (\ref{nu'=-lambda' +...}) gives the Newton's
attractive potential far from the center: $\ $%
\[
\varphi_{\text{N}}\left(  r\right)  =\frac{1}{2}c^{2}\nu\left(  r\right)
\sim-r^{-1},\qquad r\rightarrow\infty.
\]
The first term in the r.h.s. of (\ref{nu'=-lambda' +...}) appears due to the
dark matter.

The constant $a$ in the Lagrangian (\ref{L dark}) is considered the same for
any vector, be it spacelike or timelike, massive or massless. Regularity
condition for a spacelike vector (\ref{V'/(A+B)<0}) does not determine the
sign of $a.$ The negative sign $a<0$ (\ref{a < 0})\ is dictated by the
self-consistent requirement of regularity for massless and massive timelike
vectors acting together, see section \ref{Massless+massive}\ below. In view of
(\ref{V'/(A+B)<0}) and (\ref{a < 0}) the requirement of regularity for a
spacelike vector field is satisfied if $V_{0}^{\prime}$ is positive:
\[
V_{0\text{ spacelike}}^{\prime}>0.
\]
The first term with square brackets in (\ref{nu'=-lambda' +...}) with $a<0$ is
positive, both terms in the r.h.s. are of the same sign, so the presence of
dark matter increases the attraction to the center.

The curvature of space-time caused by a galaxy is small. In the linear
approximation the influence of dark and ordinary matter can be separated from
one another. For $\lambda\ll1$ (\ref{nu'=-lambda' +...}) reduces to
\begin{equation}
\nu^{\prime}=\varkappa\left\vert a\right\vert r\left[  m^{2}\left(  \phi
^{r}\right)  ^{2}+(\phi_{;M}^{M})^{2}\right]  +\frac{\lambda}{r},
\label{nu'=...+lambda/r}%
\end{equation}
where the first term does not contain $\varepsilon.$ However, the contribution
of dark matter comes from both additives.\ The vector field equation
(\ref{Field eq spherical symm}) and the Einstein equation (\ref{0-0 eq}) at
$\lambda\ll1$ are simplified:%
\begin{equation}
\left(  \phi^{r}\right)  ^{\prime\prime}+\left\{  \left[  \frac{2}%
{r}+\varkappa r\left(  \left\vert a\right\vert m^{2}\left(  \phi^{r}\right)
^{2}+\frac{1}{2}\varepsilon\right)  \right]  \phi^{r}\right\}  ^{\prime
}=-m^{2}\phi^{r} \label{vec field eq lambda very small}%
\end{equation}%
\begin{equation}
\lambda^{\prime}+\frac{\lambda}{r}=\varkappa r\left[  \left\vert a\right\vert
\left(  m^{2}\phi^{r2}-\phi_{;M}^{M}{}^{2}\right)  +\varepsilon\right]  .
\label{Eq for lambda nonrelativistic}%
\end{equation}
The boundary conditions for these equations,
\begin{equation}
\phi^{r}=\frac{1}{3}\phi_{0}^{\prime}r,\qquad\lambda=\frac{1}{3}%
\varkappa\left(  \varepsilon_{0}-\left\vert a\right\vert \phi_{0}^{\prime
2}\right)  r^{2},\qquad r\rightarrow0, \label{Boundary conditions}%
\end{equation}
are determined by the requirement of regularity in the center. Here
$\varepsilon_{0}=\varepsilon\left(  0\right)  ,$ $\phi_{0}^{\prime}=\phi
_{;M}^{M}\left(  0\right)  .$

The term $\frac{1}{2}\varkappa r\varepsilon$ in
(\ref{vec field eq lambda very small})\ reflects the interaction of dark and
ordinary matter via gravitation. If the curvature of space-time caused by the
ordinary matter is small, this term is negligible compared to $2/r.$ The
nonlinear term $\varkappa m^{2}r\left(  \phi^{r}\right)  ^{2}$ is small
compared to $2/r$ at $r\rightarrow0,$ but at $r\rightarrow\infty,$ despite of
being small, it decreases a little bit quicker than $2/r.$ Neglecting both
nonlinear terms in square brackets, the field equation
(\ref{vec field eq lambda very small}) reduces to the one in the plane
space-time. $\phi_{;M}^{M}$ obeys the Klein-Gordon equation
(\ref{eq for div fi spherical}). The regular solution is (\ref{fi^K_;K=}%
,\ref{fi^r=}):
\begin{equation}
\phi_{;M}^{M}=\phi_{0}^{\prime}\frac{\sin mr}{mr},\qquad\phi^{r}=\frac
{\phi_{0}^{\prime}}{m^{3}r^{2}}\left(  \sin mr-mr\cos mr\right)  ,
\label{analytic fi and fi;m}%
\end{equation}
where $\phi_{0}^{\prime}\equiv\phi_{;M}^{M}\left(  0\right)  .$

Substitution of (\ref{analytic fi and fi;m}) into (\ref{nu'=...+lambda/r})
results in%
\[
\nu^{\prime}\left(  r\right)  =\frac{\varkappa\left\vert a\right\vert
(\phi_{0}^{\prime})^{2}}{m^{2}r}f\left(  mr\right)  +\frac{\lambda}{r}%
,\qquad\lambda\ll1,
\]
where
\begin{equation}
f\left(  x\right)  =1-\frac{\sin2x}{x}+\frac{\sin^{2}x}{x^{2}}.
\label{f(x) analytic}%
\end{equation}

\begin{figure}
[ptbh]
\begin{center}
\includegraphics[
trim=0.000000in 0.000000in 0.000000in -0.003019in,
height=8.9095cm,
width=13.34cm
]%
{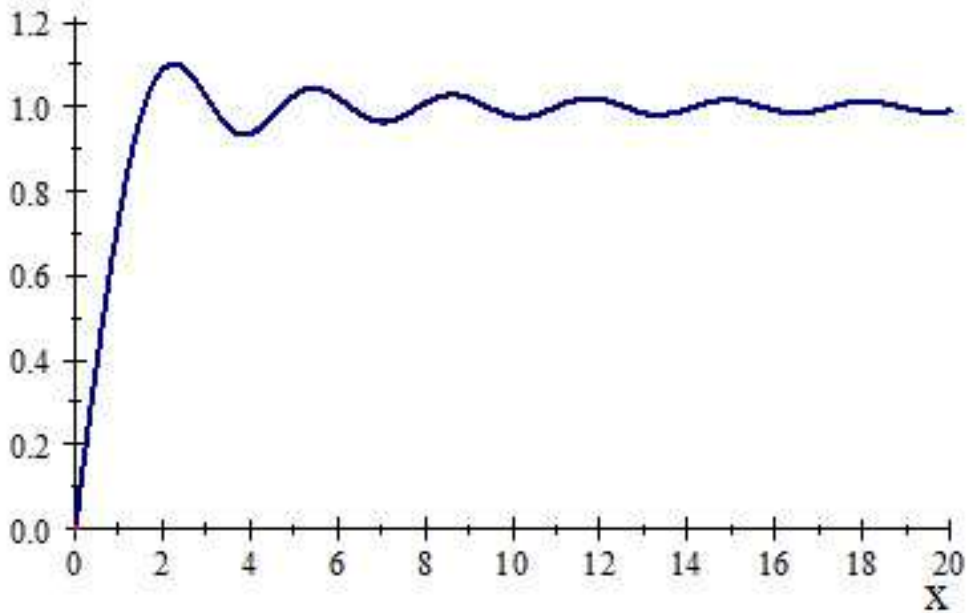}%
\caption{Universal function $\sqrt{1-\frac{\sin2x}{2x}}$ (\ref{V(r) at eps=0}%
).}%
\label{Fig4}%
\end{center}
\end{figure}

The balance of the centripetal $\frac{c^{2}\nu^{\prime}}{2}$ and centrifugal
$\frac{V^{2}}{r}$ accelerations determines the velocity $V$ of a rotating
object as a function of the radius $r$ of its orbit:%
\begin{align}
V\left(  r\right)   &  =\sqrt{V_{\text{pl}}^{2}f\left(  mr\right)
+\frac{c^{2}}{2}\lambda\left(  r\right)  },\label{V (r)}\\
V_{\text{pl}}  &  =c\sqrt{\frac{\varkappa\left\vert a\right\vert }{2}}%
\frac{\phi_{0}^{\prime}}{m}. \label{Velocity on plateau}%
\end{align}
$c$ is the velocity of light. Far from the center $\lambda\left(  r\right)  $
decreases as $1/r,$ while $f\left(  mr\right)  \rightarrow1.$ The dependence
$V\left(  r\right)  $ (\ref{V (r)}) turns at $r\gtrsim m^{-1}$ from a linear
to a plateau (\ref{Velocity on plateau}) with damping oscillations. The
plateau appears entirely due to the vector field. At the same time the vector
field contributes to $\lambda\left(  r\right)  $ as well. Substituting
(\ref{analytic fi and fi;m}) into (\ref{Eq for lambda nonrelativistic}) we get
a regular at $r\rightarrow0$ solution for $\lambda\left(  r\right)  :$%

\begin{equation}
\lambda\left(  r\right)  =2\left(  \frac{V_{\text{pl}}}{c}\right)  ^{2}\left(
\frac{\sin2mr}{2mr}-\frac{\sin^{2}mr}{m^{2}r^{2}}\right)  +\frac{\varkappa}%
{r}\int_{0}^{r}\varepsilon\left(  r\right)  r^{2}dr,\qquad\lambda
\ll1,\label{small lambda(r)}%
\end{equation}
where the last term in (\ref{small lambda(r)}) gives the Newton's potential.
Substituting (\ref{small lambda(r)}) into (\ref{V (r)}) we finally have%
\begin{equation}
V\left(  r\right)  =\sqrt{V_{\text{pl}}^{2}\left(  1-\frac{\sin2mr}%
{2mr}\right)  +c^{2}\frac{\varkappa}{2r}\int_{0}^{r}\varepsilon\left(
r\right)  r^{2}dr},\qquad\lambda\ll1.\label{Rotation curve}%
\end{equation}
Without dark matter ($\phi_{0}^{\prime}=0$) (\ref{Rotation curve}) would give
the Newton's $V\left(  r\right)  \sim1/\sqrt{r}$ at $r\rightarrow\infty.$ In
the presence of dark matter ($\phi_{0}^{\prime}\neq0$) the velocity of
rotation $V\left(  r\right)  $ tends to $V_{\text{pl}}$ at $r\rightarrow
\infty$ with damping oscillations$.$ If the contribution of stars and planets
to the total mass of a galaxy is small compared to the mass of the black hole
in the center, then outside the black hole $r\gg r_{Sch}$
\[
\varkappa\int_{0}^{r}\varepsilon\left(  r\right)  r^{2}dr=r_{Sch},
\]
$r_{Sch}$ is the Schwarzschild radius of a black hole.\ Outside a black hole
$r\gg r_{Sch},$ and the condition $\lambda\ll1$ is fulfilled for all galaxies.
Far outside the Schwarzschild radius the velocity of rotation around a black
hole is
\begin{equation}
V\left(  r\right)  =\sqrt{V_{\text{pl}}^{2}\left(  1-\frac{\sin2mr}%
{2mr}\right)  +\frac{c^{2}}{2}\frac{r_{Sch}}{r}},\qquad r\gg r_{Sch}%
.\label{V(r) at eps=0}%
\end{equation}
The deviation from the Newton's law due to the dark matter takes place at
$r\gtrsim r_{Sch}\frac{c^{2}}{V_{\text{pl}}^{2}}.$ At $r\gg r_{Sch}\frac
{c^{2}}{V_{\text{pl}}^{2}}$ the curve of rotation around a black hole is a
universal function. In dimensionless units there are no parameters, see Figure
\ref{Fig4}. \

While the contribution of dark matter to the rotation curve is described by
the universal function shown in Figure \ref{Fig4}, the distribution of stars
and planets, circulating around a black hole, differs from one galaxy to
another. The rotation curves of different galaxies look different. However,
the deviation from the Newton's $V\left(  r\right)  \sim1/\sqrt{r}$ on the
periphery of a galaxy is their common feature. Dark matter manifests itself
most clearly in the periphery of galaxies. Therefore, in order to compare with
(\ref{V(r) at eps=0}), among the numerous available rotation curves, it is
natural to choose those having the stars outside the main disk.

\begin{figure}
[ptbh]
\begin{center}
\includegraphics[
height=5.449cm,
width=14.9932cm
]%
{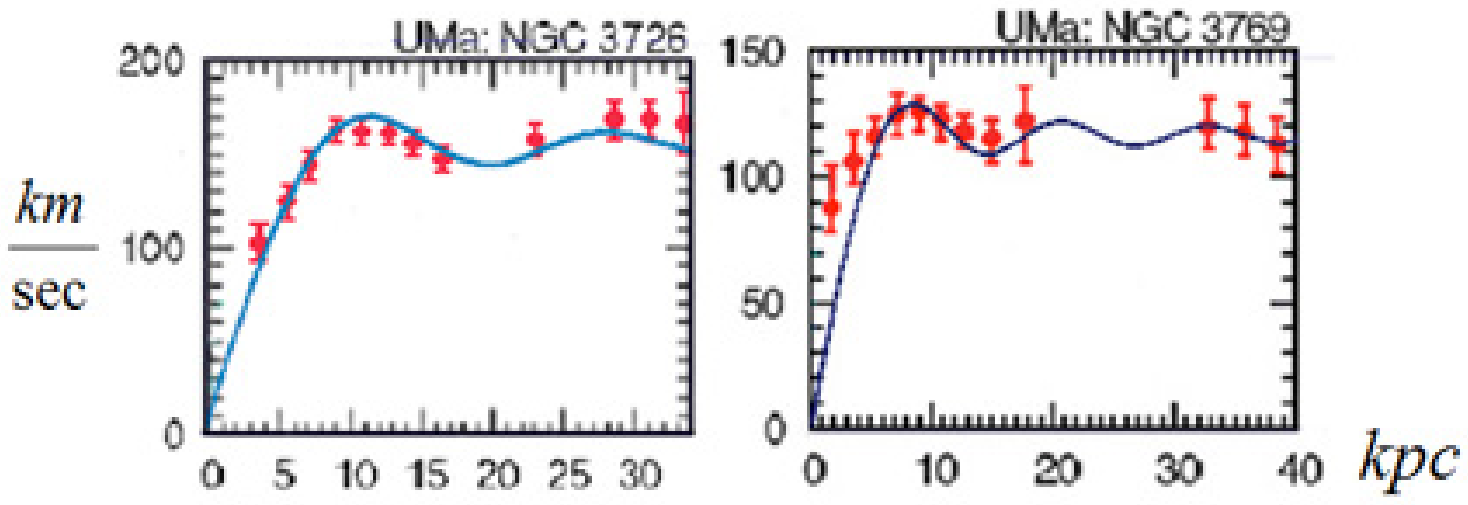}%
\caption{Rotation curves of two spiral galaxies in the Ursa Major cluster
(UMa). Abbreviation NGC stands for \textquotedblleft The New General Catalogue
of Nebulae and Clusters of Stars\textquotedblright. The vertical axis is the
velocity $V$ in $km/\sec$, and the horizontal axis is the distance $r$\ from
the center of a galaxy in $kpc$. Dots with error bars are observations. Solid
curve is fitting by $V\left(  r\right)  =V_{\text{pl}}\sqrt{1-\frac{\sin
2mr}{2mr}}$ (\ref{V(r) at eps=0}).}%
\label{Fig5}%
\end{center}
\end{figure}

Fitting the rotation curves of two such galaxies via the universal function
(\ref{V(r) at eps=0}) is presented in Figure \ref{Fig5}.
These two spiral galaxies are located in the Ursa Major cluster (UMa). Their
numbers are from \textquotedblleft The New General Catalogue of Nebulae and
Clusters of Stars\textquotedblright\ (abbreviated as NGC). It is a catalogue
of deep-sky objects in astronomy compiled by John Louis Emil Dreyer in 1888
\cite{Dreyer}, as a new version of John Herschel's Catalogue of Nebulae and
Clusters of Stars.

Agreement with the oscillations is hardly accidental. Actually, damping
oscillations of a rotation curve in the far periphery of a galaxy can be
considered as a \textquotedblleft signature of dark matter\textquotedblright.
I would strongly recommend this observational test as confirming the existence
of dark matter, along with its adequate description by a longitudinal
non-gauge vector field.

The fact of small deviations from the universal curve indicates that the main
contribution comes from the dark matter. It is in agreement with the modern
concept that there is only some 5\% of ordinary matter in the universe, while
the amount of dark matter is about five times as much, see Figure \ref{fig1}.
The fitting in Figures \ref{Fig5} also testifies that the deviation from the
central symmetry by a disk of circulating stars and planets is small. It
confirms the existence of a heavy object, like a black hole, in the center of
a galaxy. In general, the observed deviations of rotation curves from the
universal curve could clarify the average distribution of the ordinary matter
within other galaxies.

The contribution of dark matter to a rotation curve (\ref{V(r) at eps=0}) is
expressed via two observable parameters: the limiting plateau value
$V_{\text{pl}}$ and mass $m$. They allow to restore the value of the parameter
$\phi_{0}^{\prime}\equiv\phi_{;M}^{M}\left(  0\right)  $ at $r\rightarrow0$ in
the boundary conditions (\ref{Boundary conditions}): $\phi_{0}^{\prime}%
=\sqrt{2/\varkappa\left\vert a\right\vert }mV_{\text{pl }}$%
(\ref{Velocity on plateau})$.$ As far as there is no evidence of any direct
interaction of dark and ordinary matter, the origin of specific values
$\phi_{0}^{\prime}$ and $m$ of a particular galaxy depends on what happens in
the center. The values $V_{\text{pl}}$ and $m$ differ from one galaxy to
another. It looks like for each galaxy these values are driven by a heavy
object (may be a black hole, may be a neutron star) located in the center and,
by the way, supporting the central symmetry of the gravitational field.

Dark matter, described by a vector field with the Lagrangian (\ref{L dark}),
actually justifies the empirical Milgrom's hypothesis of MOND -- the modified
Newton's dynamics \cite{Milgrom}. The Newton's dynamics really gets modified
by the vector field so that the rotation curve flattens out at the far
periphery of a galaxy. This is because the perturbation of the gravitational
field due to a massive longitudinal vector field decreases slower than the
perturbation caused by the ordinary matter. The empirical Milgrom's hypothesis
of MOND was a real breakthrough in 80-ies. Naturally, basing only on the
intuitive arguments, it was scarcely possible to guess that the transition to
a plateau is accompanied by damping oscillations.

However, the question of the physical origin of dark matter remains open. In
other words, what makes $\phi_{0}^{\prime}=\phi_{;K}^{K}\left(  0\right)  $
different from zero? Solutions of the linearized Einstein equations do not
answer this question. Within the approximation of weak fields $\phi
_{0}^{\prime}$ and $m$ remain free parameters. The energy $\varepsilon\left(
r\right)  $ is an arbitrary function in the dust matter approximation, so the
parameters $\phi_{0}^{\prime}$ and $m$ of dark matter are in no way connected
with the ordinary matter.

According to the empirical MOND prediction the limiting plateau value
$V_{\text{pl MOND}}$ is connected with the mass $M$ of a galaxy:
\begin{equation}
V_{\text{pl MOND}}=\left(  \varkappa Ma_{0}\right)  ^{1/4}. \label{MOND  V(M)}%
\end{equation}
Milgrom postulates the existence of a very small acceleration $a_{0},$ and
that at $a\lesssim a_{0}$ the violation of the Newton's law takes place
\cite{Milgrom}$.$ To answer the question \textquotedblleft what the empirical
MOND relation (\ref{MOND V(M)}) should be replaced by?\textquotedblright\ one
has to find out the reason why the divergence $\phi_{;K}^{K}$ is not zero at
$r=0$. The gravitational field of a collapsing black hole is neither static,
nor weak. In the close vicinity of a black hole the velocities of circulating
stars and planets are relativistic. It is necessary to get a self-consistent
solution of the nonlinear Einstein equations. At $r\lesssim r_{Sch}$ strong
interaction via gravitation should affect the dynamic balance of the ordinary
and dark matter. For nonlinear equations the requirement of regularity can
impose an additional restriction on the parameters $\varepsilon_{0}$ and
$\phi_{0}^{\prime}$ in the boundary conditions (\ref{Boundary conditions}). It
is very likely that it will fix the connection between $\varepsilon_{0}$ and
$\phi_{0}^{\prime},$ providing the dependence of the limiting plateau value
$V_{\text{pl}}$ on the mass $M$ of a galaxy.

Actually it is a revision of equilibrium \cite{Oppenheimer Volkov} and
collapse \cite{Oppenheimer Snyder} of supermassive bodies with the dark matter
taken into account. The ordinary matter could be still considered as a
degenerate relativistic Fermi gas (see \cite{Landau-Lifshits5}, problem 3 in
the end of the paragraph 61, page 207). The dark matter should be included
into Einstein equations via the energy-momentum tensor (\ref{T_IK dark sum}%
).\ Considering a collapsing system, there is no reason to ignore a timelike
vector. It is possible that the repulsive ability of a timelike vector field
can dynamically balance the collapse. In this case there will be a regular
solution of Einstein's equations describing the internal structure of a black
hole without a singularity in the center.\ This is a worthy task for future.

One can trace two main trends in the literature in trying to unravel the
puzzle of a plateau in the rotation curves of galaxies -- to \textquotedblleft
improve\textquotedblright\ general relativity, and to compose a mixture of
fields able to fit the observations without a \textquotedblleft mysterious
dark matter\textquotedblright.

For instance, Sanders argues that \textquotedblleft..the correct theory may
well be one in which MOND reflects the influence of cosmology on local
particle dynamics and arises only in a cosmological setting\textquotedblright%
\ and concludes: \textquotedblleft It goes without saying that this theory is
not General Relativity, because in the context of General Relativity local
particle dynamics is immune to the influence of cosmology\textquotedblright%
\cite{Sanders}. Obviously, I don't share this Sanders' conclusion. I have
presented above the complete derivation from the Einstein equations
(\ref{0-0 eq}-\ref{other eqs}) to the galaxy rotation curve (\ref{V (r)}%
).\ The formulae (\ref{Rotation curve}, \ref{V(r) at eps=0}) are derived
completely within the Einstein's theory. For the time being, there is no need
in any modifications of the general relativity to explain the observable
plateau in galaxy rotation curves.

An example of opposing fields and dark matter is the Moffat's attempt of
applying a mixture of scalar, vector, and tensor fields in order
\textquotedblleft to explain the flat rotation curves of galaxies and cluster
lensing without postulating exotic dark matter\textquotedblright%
\ \cite{Moffat}. It is a question of terminology. In quantum physics there
always is a quantum particle corresponding to the field describing a material
substance. From my point of view, the fields are convenient mathematical
instruments that we utilize to describe the physical phenomena, no matter how
we name them.

According to observations the period of oscillations $\frac{\pi}{m}$ (see
Figure \ref{Fig5}) is around $15$ kpc.\ If in quantum mechanics it is the de
Broglie wavelength $\lambdabar=\frac{\hslash}{mc}$, then the rest energy of a
quantum particle, corresponding to the spacelike vector field, should be
$mc^{2}\sim\allowbreak10^{-27}$ eV. The lightest particles as candidates for
the cosmological nonbaryonic dark matter are discussed by Maxim Khlopov
\cite{Khlopov}\ in connection with spontaneously symmetry breaking in phase
transitions in the early universe.

The theory predicts the oscillating features with no baryonic counterparts in
the rotation curves of the outer regions of galaxies. As this would be the
main observational signature of existence of the dark matter, I persistently
recommend this observational test.

The \textquotedblleft unprecedented constraints on the stellar and dark matter
mass distribution within our Milky Way\textquotedblright\ are reported by Jo
Bovy and Hans-Walter Rix \cite{Jo Bovy}. Continuous progress in the accuracy
of observations would be able to provide the values of the main parameters
$\phi_{0}^{\prime}$ and $m$ for Milky Way and other galaxies and clusters.

Having the energy-momentum tensor (\ref{Tik b=c=0}), it is worth considering a
possible role of dark matter in the \textquotedblleft Pioneer
anomaly\textquotedblright\ in the scale of the solar system. It appeared that
a very small unexpected force caused an approximately constant additional
acceleration of $(8.74\pm1.33)\times10^{-10}m/s^{2}$ directed towards the Sun
for both spacecraft Pioneer 10 and Pioneer 11 \cite{Pioneer anomaly}%
,\cite{Pioneer anomaly2}. It is interesting to trace the dependence
$\nu^{\prime}\left(  r\right)  $ (\ref{nu'=...+lambda/r}) along the two
spacecraft hyperbolic orbits at distances between 20 - 70 astronomical units
(AU $=1.5\cdot10^{13}cm$) from the Sun. Manifestation of dark matter on the
periphery of the solar system would be a great surprise!

\section{\label{Cosmology}Regular cosmology}

From the standpoint of general relativity the matter curves the space-time,
giving rise to mutual attraction between the bodies. However, according to
modern observations, the Universe is expanding as a whole, despite the
gravitational attraction between material objects. The expanding solution
of\ the Einstein's equations due to the cosmological constant belongs to De
Sitter \cite{De Sitter},\cite{De Sitter2}. The expanding solutions of\ the
Einstein equations without the cosmological constant (Friedman \cite{Friedman}%
, Robertson \cite{Robertson}, Walker \cite{Walker} (FRW)) inevitably contained
the singularity. The unknown origin of expansion of the Universe, containing
only mutually attracting objects, was supposed to be hidden within the
\ singularity. For a long time the singularity was considered a general
property of the Universe. The singular point, referred to as
\textquotedblright Big Bang\textquotedblright, is still widely recognized as
the \textquotedblright date of birth\textquotedblright\ of the Universe.

Discovery of the accelerated expansion of the Universe \cite{Adam
Riess},\cite{Perlmutter} shows that the source of acceleration continues to
exist for a long time after the Big Bang.\ Naturally, the fact of accelerated
expansion gave rise to the assumption that the physical vacuum is not just the
absence of the ordinary matter. The existence of dark energy and dark matter,
as the unknown source of the Universe's expansion, is widely discussed in
modern literature and Internet \cite{wikipedia}.

The macroscopic approach to the theory of evolution of the Universe driven by
vector fields plays the central role among numerous attempts to guess the
riddle of accelerated expansion. It allows to avoid unnecessary model
assumptions (like \textquotedblleft$f\left(  R\right)  $\textquotedblright$,$
quintessence, phantom like cosmologies, ..., see a review \cite{Shin'ichi})
and remain in the classical frames of the Einstein's general relativity.
Utilization of vector fields in general relativity shows undoubtable
advantages in comparison with scalar fields and with multiplets of scalar
fields. The equations appear to be more simple, while their solutions are more
general. The solutions have additional parametric freedom, allowing to forget
the fine-tuning problem \cite{Meier1}. However, starting from the pioneer
paper by Dolgov \cite{Dolgov}, people considered mostly gauge vector fields
\cite{Rubakov Tinyakov}-\cite{Koivisto} in applications to the dark sector.
The Lorentz gauge restriction allowed to avoid the difficulty of negative
contribution to the energy. But at the same time it does not allow to utilize
all the advantages of vector fields. In general relativity (in curved
space-time) the energy is not a scalar, and its sign is not invariant against
the arbitrary coordinate transformations. Considering the vector fields in
general relativity, it is worth rejecting the gauge restriction, using instead
a more weak condition of regularity. Step by step, people are now getting rid
of the Lorentz gauge restrictions \cite{Jimenez1}-\cite{Zuntz}.

Today it is generally accepted that among the staff of the Universe only 4.5\%
is the ordinary matter, see Figure \ref{fig1} \cite{NASA sliced cake}. It is
reasonable to analyze the role of vector fields in cosmology step by step.
First step -- dark energy only (zero-mass field): about 72\%. Second step --
adding massive vector field (dark matter, about 23\%) into consideration. The
final step is to include the ordinary matter, after the main role of vector
fields is clarified.

According to observations the Universe expands, and its large scale structure
remains homogeneous and isotropic. Consider the space-time with the metric
\begin{equation}
ds^{2}=g_{IK}dx^{I}dx^{K}=(dx^{0})^{2}-e^{2F(x^{0})}\sum_{I=1}^{3}(dx^{I})^{2}
\label{Cosmological metric}%
\end{equation}
depending on only one timelike coordinate $x^{0}=ct$ \cite{FOOTNOTE}. The
metric tensor $g_{IK}$ is diagonal. The uniform and isotropic expansion is
characterized by the single metric function $F(x^{0}),$ and the rate of
expansion is $\frac{dF}{dx^{0}}\equiv F^{\prime}$. The Ricci tensor is also
diagonal:
\begin{equation}
R_{00}=-3(F^{\prime2}+F^{\prime\prime}), \label{Ricci R_00=}%
\end{equation}%
\begin{equation}
R_{II}=e^{2F}(F^{\prime\prime}+3F^{\prime2}),\qquad I>0. \label{Ricci R_11=}%
\end{equation}

\subsection{\label{Massless field as dark energy}Massless field as dark
energy}

The energy-momentum tensor (\ref{T_(0) IK= massless field}) of a massless
field acts in the Einstein equations (\ref{Einstein equations}) as a simple
addition to the cosmological constant (\ref{Lambda with tilda}):
\begin{equation}
R_{IK}-\frac{1}{2}g_{IK}R+\widetilde{\Lambda}g_{IK}=0,\quad\widetilde{\Lambda
}=\Lambda-\varkappa\left(  a\phi_{0}^{\prime\text{ }2}+V_{0}\right)  .
\label{Einstein eqs b=c=V'=0}%
\end{equation}
The contribution of the zero-mass field to the curvature of space-time remains
constant in the process of the Universe evolution. The fact that
\textquotedblleft the gauge-fixing term exactly behaves as a cosmological
constant throughout the history of the universe, irrespective of the
background evolution\textquotedblright\ had been mentioned by Jimenez and
Maroto \cite{Jimenez2}.

\begin{figure}
[ptbh]
\begin{center}
\includegraphics[
trim=0.000000in 0.000000in 0.003393in 0.000000in,
height=6.9989cm,
width=9.987cm
]%
{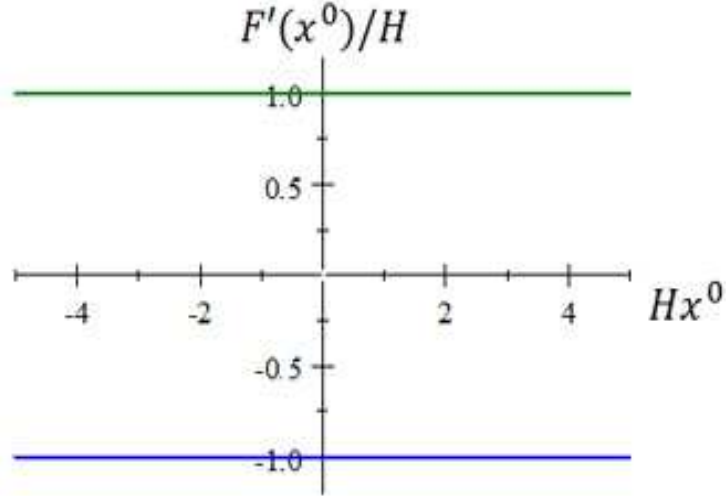}%
\caption{ de Sitter solution $F^{\prime}\left(  x^{0}\right)  /H=\pm1$. Upper
green horisontal line is expansion, and lower blue horisontal line is
compression.}%
\label{fig6}%
\end{center}
\end{figure}

The metric function
\begin{equation}
F\left(  x^{0}\right)  =\pm\sqrt{-\frac{1}{3}\widetilde{\Lambda}}(x^{0}%
-x_{0}^{0}) \label{de Sitter metric}%
\end{equation}
is the self-consistent regular solution of the Einstein equations
(\ref{Einstein eqs b=c=V'=0}), provided that
\begin{equation}
\widetilde{\Lambda}<0. \label{Lambda^<0}%
\end{equation}
$F(x^{0})$ is a linear function; $x_{0}^{0}$ is a constant of integration. The
metric\ (\ref{Cosmological metric}) with the metric function
(\ref{de Sitter metric}) is called de Sitter (or anti de Sitter, depending on
the sign definition of the Ricci tensor). It describes either expansion (sign
$+$), or contraction (sign $-$) of the Universe at a constant rate, see Figure
\ref{fig6}.%

In the case of sign $+$ the rate of expansion
\begin{equation}
F^{\prime}=H=\sqrt{-\frac{\widetilde{\Lambda}}{3}} \label{Hubble constant}%
\end{equation}
is called Hubble constant.

In general relativity the requirement of regularity for a massless field
(\ref{Lambda^<0}) replaces the artificially imposed Lorentz gauge restriction
(\ref{Lorentz gauge}) allowing to void the negative energy problem
\cite{Bogolubov Shirkov}.

The zero-mass vector field determines the constant rate of expansion.
Available today properties of the so called dark energy (presently unknown
form of matter providing the major contribution to the uniform isotropic
expansion of the Universe) can be described macroscopically by the zero-mass
vector field with a simple Lagrangian
\begin{equation}
L=a\left(  \phi_{;M}^{M}\right)  ^{2}-V_{0}. \label{Lag case a m=0}%
\end{equation}
As long as the physical nature of vacuum is not known, the \textquotedblleft
geometrical\textquotedblright\ origin of the cosmological constant $\Lambda$
and the \textquotedblleft material\textquotedblright\ contribution to
$\widetilde{\Lambda}$ (\ref{Lambda with tilda}) by the zero-mass vector field
can not be separated from one another. The combined action of the massless
field and/or the cosmological constant is described by the single parameter --
Hubble constant $H$ (\ref{Hubble constant}).

\subsection{\label{Massless+massive}Massless field + massive field}

Over the scales much larger than the distances between the galaxies the
Universe is uniform and isotropic. In the scale of the whole Universe the
massive vector field is timelike and longitudinal: the only nonzero component
is $\phi_{0},$ and it depends upon the time coordinate $x^{0}.$

The energy-momentum tensor (\ref{T_dark weak field}) for the longitudinal
massive timelike field is
\begin{align*}
T_{\text{dark }0}^{0}  &  =a(\phi_{;M}^{M})^{2}+V_{0}^{\prime}\phi_{0}^{2}\\
T_{\text{dark }I}^{K}  &  =\delta_{I}^{K}\left[  a(\phi_{;M}^{M})^{2}%
-V_{0}^{\prime}\phi_{0}^{2}\right]  ,\qquad I>0.
\end{align*}
The massless field enters the Einstein equations
\begin{equation}
3F^{\prime2}+\widetilde{\Lambda}=\varkappa a\left[  (\phi_{;M}^{M})^{2}%
+m^{2}\phi_{0}^{2}\right]  , \label{Eq  I=0 a not eq 0}%
\end{equation}%
\begin{equation}
2F^{\prime\prime}+3F^{\prime2}+\widetilde{\Lambda}=\varkappa a\left[
(\phi_{;M}^{M})^{2}-m^{2}\phi_{0}^{2}\right]  \label{Eq I>0 a not eq 0}%
\end{equation}
only via the cosmological constant $\widetilde{\Lambda}$
(\ref{Lambda with tilda}). $\widetilde{\Lambda}$ remains constant throughout
the whole history of the Universe. The massive field is described by the
function $\phi_{0}\left(  x^{0}\right)  ,$ which enters Einstein equations
(\ref{Eq I=0 a not eq 0})-(\ref{Eq I>0 a not eq 0}) directly \cite{Meier3}. In
accordance with the necessary condition of regularity for a timelike massive
vector field (\ref{V'/a >0}) $\frac{V_{0}^{\prime}}{a}>0,$ and the squared
mass of the field is $m^{2}=$ $\frac{V_{0}^{\prime}}{a}$.\ The applicability
of the equations (\ref{Eq I=0 a not eq 0})-(\ref{Eq I>0 a not eq 0}) is
restricted by the condition that the field $\phi_{0}$ is small, so that the
second and higher derivatives of the potential $V\left(  \phi_{L}\phi
^{L}\right)  $ can be ignored.

The field equations (\ref{Field eqs b=c=0}) for a longitudinal timelike field
reduce to the only one equation
\begin{equation}
\left(  \phi_{0}^{\prime}+3F^{\prime}\phi_{0}\right)  ^{\prime}+m^{2}\phi
_{0}=0, \label{Vec massive Field eq}%
\end{equation}
which is a consequence of the Einstein equations (\ref{Eq I=0 a not eq 0}%
,\ref{Eq I>0 a not eq 0}).

\subsubsection{\label{Asymptotic behavior}Asymptotic behavior at large $x^{0}%
$}

As it is confirmed below, the set (\ref{Eq I=0 a not eq 0}%
)-(\ref{Eq I>0 a not eq 0}) has regular solutions with the rate $F^{\prime
}\left(  x^{0}\right)  $ changing from $F^{\prime}\left(  -\infty\right)  =-H$
in the past to $F^{\prime}\left(  \infty\right)  =H$ in future. Far back in
the past and in the late future the temporal evolution of the massive field is
described by the equation (\ref{Vec massive Field eq}) with $F^{\prime}=\mp
H.$ Its solution
\begin{equation}
\phi_{0}\left(  x^{0}\right)  =C_{+}e^{\lambda_{+}x^{0}}+C_{-}e^{\lambda
_{-}x^{0}}, \label{psi(z) at z to inf}%
\end{equation}%
\[
\lambda_{\pm}=\left\{
\begin{array}
[c]{c}%
3H/2\pm\sqrt{\left(  3H/2\right)  ^{2}-m^{2}},\qquad x^{0}\rightarrow
-\infty,\\
-3H/2\pm\sqrt{\left(  3H/2\right)  ^{2}-m^{2}},\qquad x^{0}\rightarrow\infty,
\end{array}
\right.
\]
is a linear combination of two functions, vanishing at $x^{0}\rightarrow
\pm\infty$. The functions are monotonic if $m<$ $3H/2,$ or oscillating with a
decreasing magnitude if $m>$ $3H/2.$ If $m/H$ is small, the field decreases
very slowly:
\begin{equation}
\phi_{0}\left(  x^{0}\right)  =C_{+}\exp\left(  -\frac{2m^{2}}{9H^{2}}%
mx^{0}\right)  ,\qquad x^{0}\rightarrow\infty,\qquad\frac{m}{H}\ll1.
\label{fi(z) slow at z to inf}%
\end{equation}
In the limit $m/H\rightarrow0$ the term with $C_{-}$ disappears as
$\exp\left(  -3Hx^{0}\right)  $, while the term with $C_{+}$ becomes
indistinguishable from the massless field, which remains constant during the
whole process of evolution. In dimensional units the ratio $m/H$ is
$mc^{2}/\hbar H.$

\subsubsection{\label{Bounce}Regular bounce}

Extracting (\ref{Eq I=0 a not eq 0}) \ from (\ref{Eq I>0 a not eq 0}), we have%

\begin{equation}
F^{\prime\prime}=-a\varkappa m^{2}\phi_{0}^{2}. \label{F''=  a not equal 0}%
\end{equation}
Without massive field, i.e. if $\phi_{0}=0,$ the second derivative
$F^{\prime\prime}=0,$ and we return to the de Sitter metric with the metric
function (\ref{de Sitter metric}) describing the two isolated solutions --
compression and expansion -- at a constant rate $F^{\prime}=H=const$
(\ref{Hubble constant}).

The second order set of Einstein equations (\ref{Eq I=0 a not eq 0}%
,\ref{Eq I>0 a not eq 0}) for the unknowns $F^{\prime}$ and $\phi_{0}$ looks
more complicated than the equivalent set (\ref{Vec massive Field eq}%
,\ref{F''= a not equal 0}). At the same time the set
(\ref{Vec massive Field eq},\ref{F''= a not equal 0}) is of the third order.
Hence, it has extra solutions. So, working with the set
(\ref{Vec massive Field eq},\ref{F''= a not equal 0}), it is necessary to
eliminate extra solutions that are not the solutions of the Einstein equations
(\ref{Eq I=0 a not eq 0},\ref{Eq I>0 a not eq 0}).

The time coordinate $x^{0}$ is a cyclic variable, and it is convenient to set
the origin $x^{0}=0$ at a moment when $F^{\prime}=0.$ Initial conditions for
the Einstein equations contain only $\phi_{0}\left(  0\right)  .$ The
derivative $\phi_{0}^{\prime}\left(  0\right)  $ is strictly fixed by the
solutions of (\ref{Eq I=0 a not eq 0},\ref{Eq I>0 a not eq 0}). As for the set
(\ref{Vec massive Field eq},\ref{F''= a not equal 0}), the value $\phi
_{0}^{\prime}\left(  0\right)  $ in the initial conditions is a free
parameter, independent of $\phi_{0}\left(  0\right)  .$ The connection between
$\phi_{0}\left(  0\right)  $ and $\phi_{0}^{\prime}\left(  0\right)  ,$
eliminating extra solutions,\ follows from the equation
(\ref{Eq I=0 a not eq 0}) at $x^{0}=0$:
\begin{equation}
\phi_{0}^{\prime2}\left(  0\right)  +m^{2}\phi_{0}^{2}\left(  0\right)
=\frac{\widetilde{\Lambda}}{\varkappa a},\qquad F^{\prime}\left(  0\right)
=0. \label{Initial conditions}%
\end{equation}
The l.h.s. of (\ref{Initial conditions}) is positive. The initial conditions
(\ref{Initial conditions}) are self-consistent if both $\widetilde{\Lambda}$
\ and $a$ \ are of the same sign. According to the requirement of regularity
of the de Sitter metric (\ref{Lambda^<0}) $\widetilde{\Lambda}$ is negative.
Hence $a$ is negative too:%
\begin{equation}
a<0. \label{a < 0}%
\end{equation}
Then $F^{\prime\prime}$ (\ref{F''= a not equal 0}) is positive, $F^{\prime
\prime}>0.$ We conclude, that the massive timelike vector field makes the rate
of evolution $F^{\prime}\left(  x^{0}\right)  $ a monotonically growing
function from $-H$ in the past to $+H$ in future. The universe contracts at
$x^{0}<0,$ and expands at $x^{0}>0.$ $x^{0}=0$ is the moment of maximum compression.

One of the two constants $\phi_{0}$ and $\phi_{0}^{\prime}$ at $x^{0}=0$
remains arbitrary within the initial conditions (\ref{Initial conditions}).
The Einstein equations (\ref{Eq I=0 a not eq 0},\ref{Eq I>0 a not eq 0}) are
$x^{0}\rightarrow-x^{0}$ invariant. In the case $\phi_{0}^{\prime}\left(
0\right)  =0$ the field $\phi_{0}\left(  x^{0}\right)  $ is a symmetric
function, and if $\phi_{0}\left(  0\right)  =0$ it is an antisymmetric one. In
both cases $F^{\prime}\left(  x^{0}\right)  $ is antisymmetric. If both
constants $\phi_{0}$ and $\phi_{0}^{\prime}$ at $x^{0}=0$\ are not zeroes, a
regular solution still exists, but there is no symmetry with respect to
$x^{0}\rightarrow-x^{0}.$ The scale factor $R=e^{F}$ decreases with time while
$F^{\prime}<0,$ reaches its minimum, and grows when $F^{\prime}$ becomes positive.

\begin{figure}
[ptbh]
\begin{center}
\includegraphics[
trim=0.000000in 0.000000in 0.003393in 0.000000in,
height=6.7543cm,
width=9.987cm
]%
{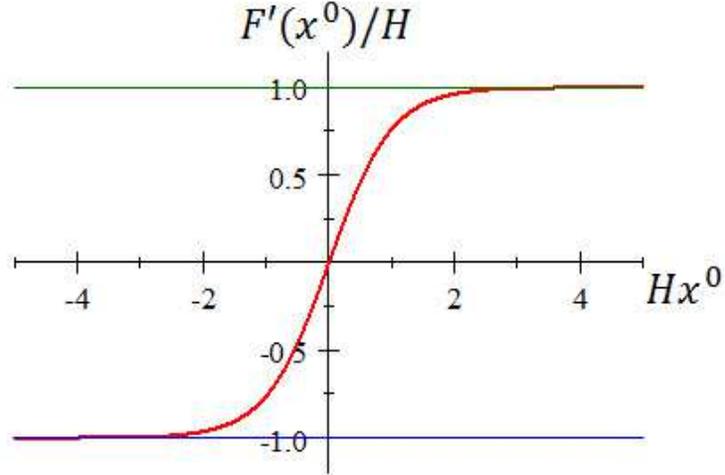}%
\caption{Compression-to-expansion transition. Red line is the rate of
evolution $F^{\prime}\left(  x^{0}\right)  /H=\tanh\left(  3Hx^{0}\right)  $
(\ref{F'(x^0)=H tanh()}). }%
\label{fig7}%
\end{center}
\end{figure}

In the case of a small mass,
\begin{equation}
m\ll H,\label{m << H}%
\end{equation}
(in dimensional units $mc^{2}\ll\hbar H$) the compression-to-expansion
transition is described by the analytical solution for the symmetric
configuration as follows:
\begin{align}
F^{\prime}\left(  x^{0}\right)   &  =H\tanh\left(  3Hx^{0}\right)
,\label{F'(x^0)=H tanh()}\\
\phi_{0}\left(  x^{0}\right)   &  =\sqrt{\frac{\widetilde{\Lambda}}{\varkappa
a}}\frac{1}{m\cosh\left(  3Hx^{0}\right)  },\quad m\ll H.\label{fio(x^0) m<<H}%
\end{align}
The rate of evolution $F^{\prime}\left(  x^{0}\right)  $
(\ref{F'(x^0)=H tanh()}) is shown in Figure \ref{fig7}.%
With no ordinary matter the time interval of transition is of the order of
Hubble time $\sim1/3H.$ At $m\ll H$\ it does not depend on the mass $m$ of the
massive field. The scale factor is
\[
R\left(  x^{0}\right)  =e^{F\left(  x^{0}\right)  }=e^{F_{0}}\left[
\cosh\left(  3Hx^{0}\right)  \right]  ^{1/3},\quad m\ll H.
\]
The metric function enters the Einstein equations (\ref{Eq I=0 a not eq 0}%
,\ref{Eq I>0 a not eq 0}) only via the derivatives of $F\left(  x^{0}\right)
,$ but not directly. Without ordinary matter $F_{0}=F\left(  0\right)  $ is a
free parameter, and $R\left(  x^{0}\right)  $ is defined up to an arbitrary
constant factor $e^{F_{0}}$.\ The acceleration $F^{\prime\prime}$ is
positive:
\begin{equation}
F^{\prime\prime}\left(  x^{0}\right)  =\frac{3H^{2}}{\cosh^{2}\left(
3Hx^{0}\right)  }>0,\quad m\ll H.\label{F''(x^o)=   m<<H}%
\end{equation}

Like an elastic spring,\ the longitudinal vector field enables the transition
from compression to expansion. The kinetic energy of contraction completely
converts at $x^{0}=0$ into potential energy of the compressed vector field,
and at $x^{0}>0$ the energy is being released back in the form of the kinetic
energy of expansion.

A timelike longitudinal massive vector field displays repulsive elasticity. It
can hardly be attributed to a particle of ordinary matter, because there is no
reference frame where such particle could be at rest. However, a timelike
vector field can be associated with a topological defect, inevitably arising
in a phase transition with spontaneous symmetry breaking \cite{Meierovich2}%
,\cite{Khlopov}. It is worth mentioning that in the case (\ref{m << H}) in the
state of maximum compression the field $\phi_{0}\left(  0\right)  $
(\ref{fio(x^0) m<<H}) is proportional to $H/m\gg1.$ Depending on the
parameters of the potential $V\left(  \phi^{M}\phi_{M}\right)  $ for very
small $m$ the field $\phi_{0}\left(  0\right)  $ can be too big to omit the
second and higher derivatives of $V\left(  \phi^{M}\phi_{M}\right)  .$ Then in
the state of maximum compression a phase transition with spontaneous symmetry
breaking can take place. The idea to consider the topological defect as aether
looks nice. Spontaneous breaking of Lorentz symmetry caused by a timelike
vector field is a subject of research entitled "Einstein-aether model", see
\cite{Haghani}-\cite{Armendariz-Picon et al},\cite{Armendariz-Picon2} and
references there in.

In the opposite case of a large mass,%
\begin{equation}
m\gg H, \label{mu >> 1}%
\end{equation}
the field $\phi_{0}\left(  x^{0}\right)  $ is a rapidly oscillating function
as compared with $\overline{F^{\text{ }\prime}}\left(  x^{0}\right)  .$ The
solution is%
\begin{equation}
\overline{F^{\text{ }\prime}}\left(  x^{0}\right)  =H\tanh\left(  \frac{3}%
{2}Hx^{0}\right)  , \label{overline(F)_(z) =}%
\end{equation}%
\begin{equation}
\phi_{0}\left(  x^{0}\right)  =\sqrt{\frac{\widetilde{\Lambda}}{\varkappa a}%
}\frac{\cos\left(  mx^{0}+\varphi\right)  }{m\cosh\left(  \frac{3}{2}%
Hx^{0}\right)  },\quad m\gg H. \label{fi(z) = mu>>1}%
\end{equation}
$\overline{F^{\text{ }\prime}}\left(  x^{0}\right)  $ is the rate of
expansion, averaged over the rapid oscillations.

\begin{figure}
[ptbh]
\begin{center}
\includegraphics[
trim=0.000000in 0.000000in -0.001798in -0.001089in,
height=7.2435cm,
width=9.5463cm
]%
{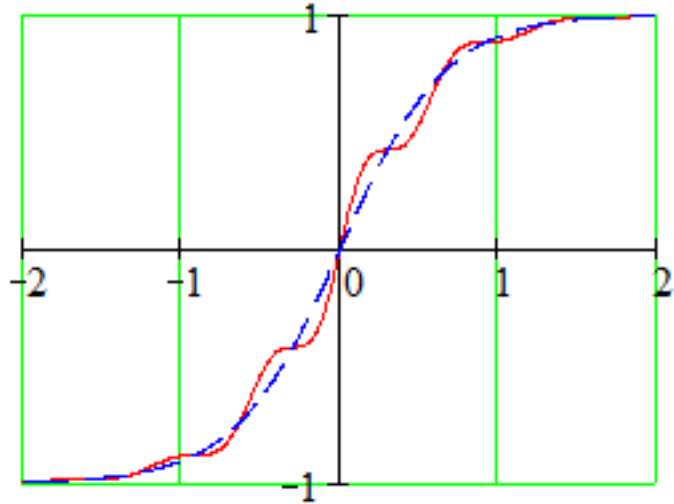}%
\caption{The rate of evolution in the case $m\gg H.$ $\ F^{\prime}\left(
x^{0}\right)  /H-$ red solid curve, found numerically for $m/H=5,$
$\phi^{\prime}\left(  0\right)  =0.$ \ The averaged rate $\overline{F^{\text{
}\prime}}\left(  x^{0}\right)  /H=\tanh\left(  \frac{3}{2}Hx^{0}\right)  -$
blue dashed line.}%
\label{fig8}%
\end{center}
\end{figure}

Oscillations of $\phi_{0}\left(  x^{0}\right)  $\ at large $m$ initiate weak
vibrations of the rate $F^{\prime}\left(  x^{0}\right)  $ around the averaged
value $\overline{F^{\text{ }\prime}}\left(  x^{0}\right)  ,$ see Figure
\ref{fig8}.
Red curve is the numerical solution for $m/H=5,$ $\phi^{\prime}\left(
0\right)  =0.$ Blue dashed line -- analytical solution $\overline{F^{\text{
}\prime}}\left(  x^{0}\right)  $ (\ref{overline(F)_(z) =}). The phase
$\varphi\ $ depends on the relation between the initial values $\phi
_{0}\left(  0\right)  $ and $\phi_{0}^{\prime}\left(  0\right)  .$%

\begin{figure}
[ptbh]
\begin{center}
\includegraphics[
height=6.5076cm,
width=15.5162cm
]%
{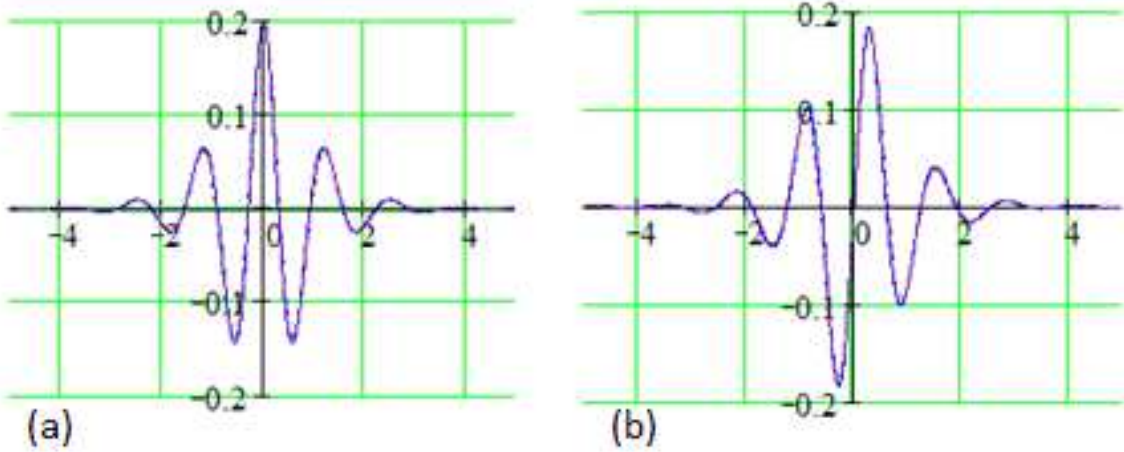}%
\caption{Symmetric $\varphi=0$ (a), and antisymmetric $\varphi=-\pi/2$ (b)
time-like massive vector fields $\sqrt{\frac{\varkappa a}{\widetilde{\Lambda}%
}}\phi_{0}\left(  x^{0}\right)  $ (\ref{fi(z) = mu>>1})$,$ $m/H=5$. }%
\label{fig9}%
\end{center}
\end{figure}

In Figures \ref{fig9}a and \ref{fig9}b $\phi_{0}\left(  x^{0}\right)  ,$ found
numerically for $m/H=5$, practically coincide with found analytically
(\ref{fi(z) = mu>>1}). The initial condition $\phi_{0}^{\prime}\left(
0\right)  =0$ for the symmetric solution in Figure \ref{fig9}a corresponds to
$\varphi=0$ in (\ref{fi(z) = mu>>1}). The antisymmetric solution in Figure
\ref{fig9}b (the initial condition $\phi_{0}\left(  0\right)  =0$) coincides
with (\ref{fi(z) = mu>>1}) at $\varphi=-\pi/2.$

See \cite{Meierovich1} for the details of\ analytical and numerical solutions.

\subsubsection{\label{Acceleration or coefficient?}Acceleration and
\textquotedblleft acceleration coefficient\textquotedblright}

In modern literature the uniform and isotropic evolution of the Universe is
described by the rate of expansion (Hubble parameter) $H\left(  x^{0}\right)
=R^{\prime}/R=F^{\prime}\left(  x^{0}\right)  $ and by the artificially
introduced so-called "acceleration coefficient" $q,$
\begin{equation}
q=R^{\prime\prime}/R=F^{\prime\prime}+F^{\prime2}. \label{def q}%
\end{equation}
$R\left(  x^{0}\right)  =e^{F\left(  x^{0}\right)  }$ is the scale factor.
This way defined $q$ remains positive even if the evolution goes at a constant
rate $F^{\prime}=H=const$ (as under the action of the massless field):
\begin{equation}
q=H^{2}>0,\quad F^{\prime\prime}=0. \label{q>0}%
\end{equation}
The commonly spread statement, that the dark energy is the source of
accelerated expansion, is based on the definition (\ref{def q}). It is worth
clarifying that the term "acceleration" \ is used here for $F^{\prime\prime}$
-- the velocity of variation of the rate $F^{\prime}:$ \ $F^{\prime\prime
}=\left(  R^{\prime}/R\right)  ^{\prime}$. Not to be confused with the
\textquotedblleft acceleration coefficient\textquotedblright\ $q=R^{\prime
\prime}/R$.

\subsection{\label{Scenarios of evolution}Scenarios of evolution of the
universe driven by vector fields and ordinary matter}

The above analysis of the equations (\ref{Vec massive Field eq}%
,\ref{F''= a not equal 0})\ facilitates clarifying the solutions of the
Einstein equations with the ordinary matter taken into account.

Applying the general relativity to the Universe as a whole it is natural to
consider the ordinary matter (stars, galaxies, ...) as separate subsystems
located at large distances from one another. Averaged over the distances much
larger than the distance between the objects, the ordinary matter can be
considered macroscopically as a uniformly distributed dust. In the dust matter
approximation all components of the macroscopic energy-momentum tensor
(\ref{Tom_IK}), except $T_{00}=\varepsilon g_{00},$ are zeros. In the
cosmological metric (\ref{Cosmological metric}) $g_{00}=1$.

In the process of uniform and isotropic evolution the average energy density
of matter $\varepsilon$ depends only on $x^{0}$. Due to Bianchi identities the
covariant divergence of the energy-momentum tensor is zero: $T_{I;K}%
^{K}=e^{-3F}\frac{d\left(  e^{3F}\varepsilon\right)  }{dx^{0}}\delta_{I0}=0.$
Thus $\varepsilon e^{3F}=const$, the energy density of the ordinary matter is
inverse proportional to the cube of scale factor $R=e^{F}.$ If we accept that
$\varepsilon\left(  x^{0\ast}\right)  =\varepsilon_{0\text{ }}$is the averaged
energy density of the ordinary matter now, then $\varepsilon\left(
x^{0}\right)  =\varepsilon_{0}e^{-3F\left(  x^{0}\right)  },$ and the present
moment $x^{0\ast}$ is defined by
\begin{equation}
F\left(  x^{0\ast}\right)  =0. \label{F(x^0*)=0}%
\end{equation}
In the process of expansion the metric function is negative in the past:
$F\left(  x^{0}\right)  <0$ at $x\,^{0}<x^{0\ast}$.

\subsubsection{\label{Einstein equations 2}Einstein equations and initial
conditions}

With the ordinary dust matter taken into account, instead of the Einstein
equation (\ref{Eq I=0 a not eq 0})\ we have
\begin{equation}
3F^{\prime2}+\widetilde{\Lambda}=\varkappa a\left[  (\phi_{;M}^{M})^{2}%
+m^{2}\phi_{0}^{2}\right]  +\varkappa\varepsilon_{0}e^{-3F},\quad I=0.
\label{Eq I=0}%
\end{equation}
The second equation (\ref{Eq I>0 a not eq 0}) remains the same. The vector
field equation (\ref{Vec massive Field eq}) also remains the same, and the
equation (\ref{F''= a not equal 0}) is changed to
\begin{equation}
F^{\prime\prime}=\varkappa|a|m^{2}\phi_{0}^{2}-\frac{1}{2}\varkappa
\varepsilon_{0}e^{-3F}. \label{F_,z,z = ...}%
\end{equation}
The equation (\ref{F_,z,z = ...}) resembles the Newton's law: acceleration
$F^{\prime\prime}$ is proportional to the \textquotedblleft repulsing
force\textquotedblright\ $\varkappa|a|m^{2}\phi_{0}^{2}$ minus the
\textquotedblleft attracting force\textquotedblright\ $\frac{1}{2}%
\varkappa\varepsilon_{0}e^{-3F}.$\ The equations (\ref{Vec massive Field eq}%
,\ref{F_,z,z = ...}) and the initial conditions
\begin{equation}
\frac{\varkappa a}{\widetilde{\Lambda}}\left[  \phi_{0}^{\prime2}\left(
0\right)  +m^{2}\phi_{0}^{2}\left(  0\right)  \right]  =1+\Omega e^{-3F_{0}%
},\quad F^{\prime}\left(  0\right)  =0,\quad F\left(  0\right)  =F_{0}%
,\quad\widetilde{\Lambda}<0,\text{ \ }a<0, \label{bound cond with ord matter}%
\end{equation}
following from (\ref{Eq I=0}) at $F^{\prime}=0,$\ contain five dimensionless
parameters: $\frac{m}{H},\Omega,F_{0},\sqrt{\frac{\varkappa a}{\widetilde
{\Lambda}}}\phi_{0}\left(  0\right)  ,$ and $\sqrt{\frac{\varkappa
a}{\widetilde{\Lambda}}}\phi_{0}^{\prime}\left(  0\right)  $. In view of the
connection (\ref{bound cond with ord matter}) four of them are independent. As
usual, the parameter $\Omega,$
\begin{equation}
\Omega=-\frac{\varkappa\varepsilon_{0}}{\widetilde{\Lambda}}=\frac
{\varkappa\varepsilon_{0}}{3H^{2}}, \label{Omega}%
\end{equation}
denotes the ratio of today's energy density of the ordinary matter to the
density of kinetic energy of expansion at a constant rate $H$. According to
the NASA's \textquotedblleft sliced cake\textquotedblright\ diagram (Figure
\ref{fig1})
\begin{equation}
\Omega\sim0.06. \label{Omega 0.06}%
\end{equation}
Actually the exact value of $\Omega$\ becomes important in the vicinity of
$\Omega e^{-3F_{0}}=1:$\ regular oscillating solutions appear at $\Omega
e^{-3F_{0}}>1$ in addition to the cosmological ones, see Section
\ref{Regular oscillating solutions} below.\ 

\subsubsection{\label{Regular cosmological solutions}Regular cosmological
solutions}

Equations (\ref{Vec massive Field eq},\ref{F_,z,z = ...}) with initial
conditions (\ref{bound cond with ord matter}) are easily integrated
numerically. Regular solutions are free from any fine tuning. Moreover, the
existing parametric freedom leads to a great variety of possible regular
scenarios of evolution.

Numerical analysis (see details in \cite{Meierovich1}) shows, that if both
$\phi_{0}\left(  0\right)  ,$ and $\phi_{0}^{\prime}\left(  0\right)  $ are
not zeros at a turning point $F^{\prime}\left(  0\right)  =0,$ then there can
be other even more sharp turning points with $\phi_{0}^{\prime}$ more close to
zero. With the ordinary matter taken into account the metric function
$F\left(  x^{0}\right)  $ enters the equation (\ref{F_,z,z = ...}) directly.
The parameter $F_{0}=F\left(  0\right)  <0$ determines the degree of maximum
compression at the turning point $F^{\prime}\left(  0\right)  =0.$ The peak
value of the rate of expansion grows exponentially with the increasing
negative value of $F_{0},$ $F^{\prime}\sim e^{-3F_{0}},$ while the width of
the transition decreases exponentially. It resembles inflation, except that
there is no singularity. The regular contraction-to-expansion transition is
often referred to as \textquotedblleft nonsingular bounce\textquotedblright%
\ \cite{Creminelli},\cite{Lin},\cite{Steinhardt}. In the literature there are
attempts to construct a self-consistent model in order to explain from a
unified viewpoint the inflation in the early Universe and the late-time
accelerated expansion \cite{Balakin}. However, one should keep in mind that
the dust matter approximation is not applicable until the galaxies become
non-interacting systems located at far distances from one another.

In the most interesting case of small $m$ (\ref{m << H}) the transition from
contraction to expansion, resembling inflation, can be described analytically
\cite{Meierovich1}. Utilizing the fact that at $m\rightarrow0$ the
antisymmetric term acts as a cosmological constant, it is natural to consider
its contribution as already included into $\widetilde{\Lambda}$
(\ref{Einstein eqs b=c=V'=0}), so that at $m\rightarrow0$ $\ \ \widetilde
{\Lambda}$ corresponds to the observable Hubble constant $H$
(\ref{Hubble constant}), and $\phi_{0}^{\prime}\left(  0\right)  =0$ in the
initial conditions (\ref{bound cond with ord matter}). At $m\ll H$ one can
neglect the term $m^{2}\phi_{0}$ in the field equation
(\ref{Vec massive Field eq}) and express $\phi_{0}\left(  x^{0}\right)  $ via
$F\left(  x^{0}\right)  $ $:$%
\begin{equation}
\phi_{0}\left(  x^{0}\right)  =\phi_{0}\left(  0\right)  \exp\left\{
-3\left[  F\left(  x^{0}\right)  -F_{0}\right]  \right\}  ,\qquad m\ll H.
\label{fi(z) symm mu<<1}%
\end{equation}
\ Substituting (\ref{fi(z) symm mu<<1}) into (\ref{F_,z,z = ...}), we exclude
the field $\phi_{0}$ and come to the single equation for $F:$%
\[
F^{\prime\prime}=3H^{2}\left[  \left(  e^{6F_{0}}+\Omega e^{3F_{0}}\right)
e^{-6F}-\frac{1}{2}\Omega e^{-3F}\right]  ,\qquad m\ll H.
\]
Its regular solution with initial conditions $F^{\prime}\left(  0\right)
=0,\quad F\left(  0\right)  =F_{0}$ is%
\begin{equation}
F\left(  x^{0}\right)  =F_{0}+\frac{1}{3}\ln\left[  \left(  1+\frac{\Omega}%
{2}e^{-3F_{0}}\right)  \cosh\left(  3Hx^{0}\right)  -\frac{\Omega}%
{2}e^{-3F_{0}}\right]  ,\qquad m\ll H. \label{F(z)=  mu<<1}%
\end{equation}
In the limit $m\ll H$ the metric function $F\left(  x^{0}\right)  $ does not
depend on the mass $m$ of the vector field. $F_{0}=F\left(  0\right)  <0$
remains a free parameter. It determines the strength of\ maximum compression
at $x^{0}=0.$

\begin{figure}
[ptbh]
\begin{center}
\includegraphics[
trim=0.000000in 0.000000in 0.002072in 0.001357in,
height=6.0205cm,
width=9.1541cm
]%
{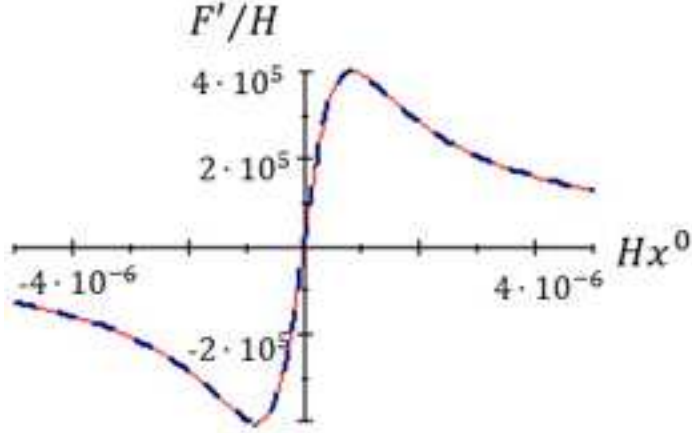}%
\caption{$F^{\prime}/H$ in the vicinity of the turning point. Blue dashed
curve is the numerical result for $F_{0}=-10,$ $m/H=10,$ $\Omega=0.06$. It
coinsides with the red solid curve -- analytical solution
(\ref{F_,z(z)= mu << 1}).}%
\label{fig10}%
\end{center}
\end{figure}

For the rate of evolution $F^{\prime}\left(  x^{0}\right)  ,$ and\ for the
scale factor $R\left(  x^{0}\right)  $ we \ get
\begin{align}
F^{\prime}\left(  x^{0}\right)   &  =H\frac{\sinh\left(  3Hx^{0}\right)
}{\cosh\left(  3Hx^{0}\right)  -\left(  1+\frac{2}{\Omega}e^{3F_{0}}\right)
^{-1}},\label{F_,z(z)=   mu << 1}\\
R\left(  x^{0}\right)   &  =\left[  \left(  e^{3F_{0}}+\frac{1}{2}%
\Omega\right)  \cosh\left(  3Hx^{0}\right)  -\frac{1}{2}\Omega\right]
^{\frac{1}{3}},\qquad m\ll H.\label{R(z)=   mu << 1}%
\end{align}
Analytical solutions (\ref{F(z)= mu<<1}-\ref{R(z)= mu << 1}), derived for
$m\ll H,$ are as well applicable for $m\sim H$ in the vicinity of the turning
point, if $\left\vert F_{0}\right\vert \gg1$. See Figure \ref{fig10}, where
the variation of $F^{\prime}\left(  x^{0}\right)  $, found numerically for
$\Omega=0.06,$ $m/H=10,$ and $F_{0}=-10,$ practically coincides with
(\ref{F_,z(z)= mu << 1}) in the vicinity of the turning point $x^{0}=0$.%
It is because for very large negative $F_{0}$ the width of the
contraction-to-expansion transition $\Delta x^{0}$ is very narrow:
\[
H\Delta x^{0}\sim\frac{2}{3\sqrt{\Omega}}e^{-3|F_{0}|/2},\quad F_{0}%
<0,\quad\left\vert F_{0}\right\vert \gg1.
\]

In the process of compression the repulsing term $\sim e^{-6F}$ in
(\ref{F_,z,z = ...}) increases faster than the compressing term $\sim$
$e^{-3F}$. It is the reason why a regular bounce replaces the singularity
independently of how big the negative $F_{0}$ is. After the bounce the
repulsing term decreases faster than the compressing one, leading to matter
domination over the field at late times.\ 

The limits of applicability of Eq. (\ref{F_,z,z = ...}) are connected with the
dust matter approximation and with omitting the second and higher derivatives
of the potential $V\left(  \phi^{K}\phi_{K}\right)  $. The symmetric field at
the bounce $\phi^{2}\left(  0\right)  \sim\varkappa\varepsilon_{0}e^{-6F_{0}%
}/m^{2}$ can be very large, and it should lead to a phase transition with
spontaneous symmetry breaking. Naturally, in this case the solution
(\ref{F_,z(z)= mu << 1}) would be unstable. The analysis of possible symmetry
breaking in the limit $m/H\ll1$ deserves a separate consideration.

\begin{figure}
[ptbh]
\begin{center}
\includegraphics[
trim=0.000000in 0.000000in 0.001724in 0.001503in,
height=7.9774cm,
width=15.1978cm
]%
{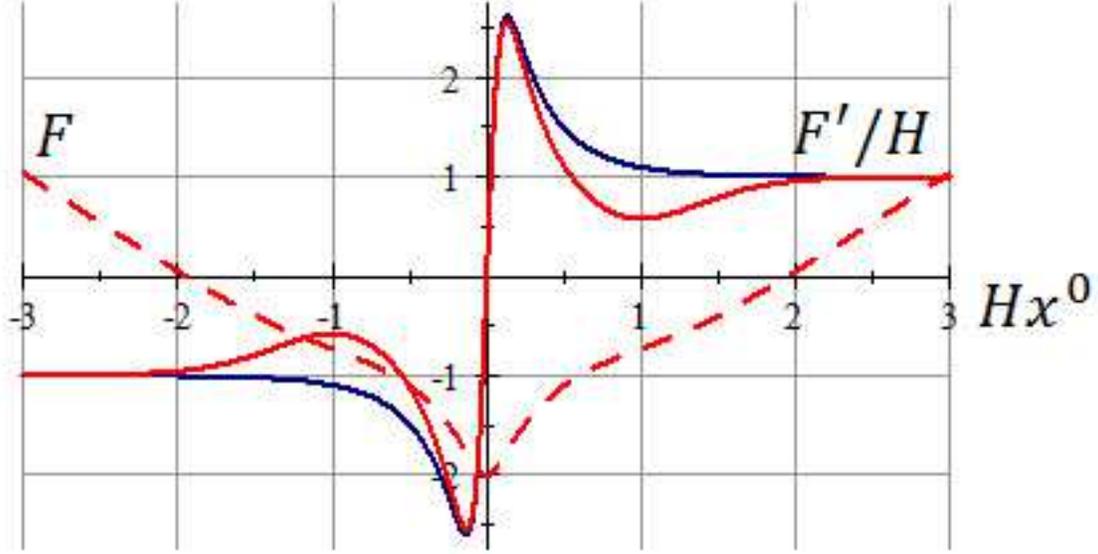}%
\caption{Solid curves are the rates of evolution $F^{\prime}/H$ for
$\Omega=0.06,$ $F_{0}=-2.$ Blue curve is the analytical solution
(\ref{F_,z(z)= mu << 1}) for $m/H\ll1,$ and the red solid curve is the
numerical solution for $m/H=1.$ The dashed red curve is the metric function
$F$. Horizontal axis is the dimensionless time $Hx^{0}.$ In accordance with
the \textquotedblleft calendar\textquotedblright\ (\ref{F(x^0*)=0}) today's
date is $Hx^{0\ast}\approx2.$ The moment of switching from deceleration to
acceleration ($F^{\prime\prime}=0$) is $Hx^{0}\approx1.$}%
\label{fig11}%
\end{center}
\end{figure}

According to the analysis of the Hubble space telescope data \cite{Suzuki},
the expansion of the Universe switched from \ deceleration to acceleration at
about a half of the age of the Universe. In the analytical solution
(\ref{F_,z(z)= mu << 1}) the second derivative $F^{\prime\prime}\left(
x^{0}\right)  $ is negative on the slope of $F^{\prime}\left(  x^{0}\right)  $
after the peak (blue curve in Figure \ref{fig11}). The expansion continues
($F^{\prime}>0$), but in the case $m\ll H$ it goes with deceleration
($F^{\prime\prime}<0$) all the time after the peak.

The red solid curve in Figure \ref{fig11} is the numerical solution for
$m/H=1,$ $F_{0}=-2,$ $\Omega=0.06.$ In the case $m/H=1$\ the expansion
switches from deceleration to acceleration at about a half time passed from
the turning point $x^{0}=0$ to the present moment $Hx^{0\ast}\approx2$ (where
$F(x^{0\ast})=0,$ see the \textquotedblleft calendar\textquotedblright%
\ (\ref{F(x^0*)=0})). It is in accordance with the Hubble space telescope data
\cite{Suzuki}. The particular value $F_{0}=-2$ in Figure \ref{fig11} is taken
not big for better clarity. The peak value of $F^{\prime}$ grows exponentially
with increasing $|F_{0}|.$ However at a fixed value $m/H\gtrsim1$ the
qualitative picture remains the same: the transition from deceleration to
acceleration does not disappear.

\subsubsection{\label{Regular oscillating solutions}Regular oscillating
solutions (with positive $\widetilde{\Lambda}$)}

\begin{figure}
[ptbh]
\begin{center}
\includegraphics[
height=5.0251cm,
width=13.8967cm
]%
{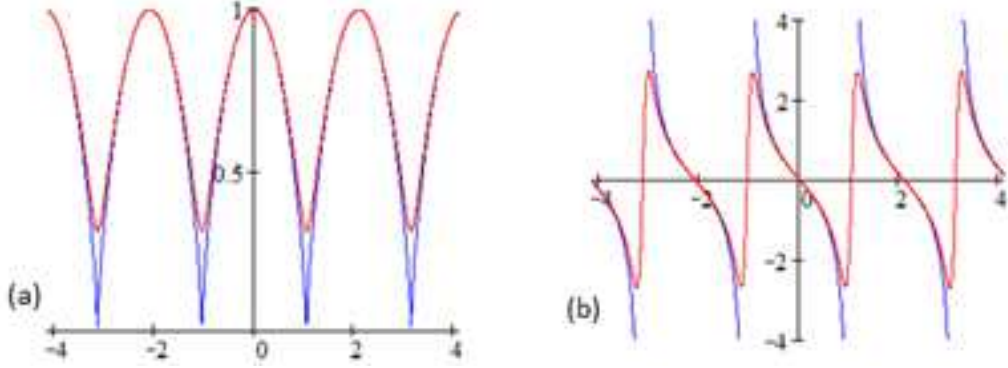}%
\caption{ Scale factor $\frac{R}{R\left(  0\right)  }$ (a) and rate of
evolution $\frac{F^{\prime}}{H}$ (b). The horizontal axis is \textquotedblleft
time\textquotedblright\ $Hx^{0}$. Red curves -- regular numerical solution
(coinsiding with (\ref{R oscil}) and (\ref{F_,z oscil})) for $m/H=0.02,$
$\Omega\exp\left(  -3F_{0}\right)  =1.032.$ Blue curves with periodic
singularity are the fine tuned solutions (\ref{R(z) no field}) \ and
(\ref{F_,z (z) no field}) at the lower boundary of the area of regular
oscillating solutions (with $\widetilde{\Lambda}>0)$.}%
\label{fig12}%
\end{center}
\end{figure}

There is an important difference between the initial conditions
(\ref{Initial conditions}) and (\ref{bound cond with ord matter}). The
relation (\ref{Initial conditions}) can be satisfied only if $\widetilde
{\Lambda}<0,$ provided that $a<0.$ Appearance of the term $\Omega e^{-3F_{0}}$
in (\ref{bound cond with ord matter}) admits the solutions with positive
$\widetilde{\Lambda}.$ If $\widetilde{\Lambda}$ changes sign, then $H$
(\ref{Hubble constant}) becomes imaginary. The equations
(\ref{Vec massive Field eq},\ref{F_,z,z = ...}) are invariant against
$H\rightarrow iH,$ but the initial conditions
(\ref{bound cond with ord matter}) are not:
\begin{equation}
\frac{\varkappa|a|}{\widetilde{\Lambda}}\left[  \phi_{0}^{\prime2}\left(
0\right)  +m^{2}\phi_{0}^{2}\left(  0\right)  \right]  =-1+\Omega e^{-3F_{0}%
},\quad F^{\prime}\left(  0\right)  =0,\quad F\left(  0\right)  =F_{0}%
,\quad\widetilde{\Lambda}>0.
\label{bound cond with ord matter Lambda positive}%
\end{equation}
A necessary condition for regular solutions with $\widetilde{\Lambda}>0$ is
the existence of an extremum moment $\left(  F^{\prime}\left(  0\right)
=0\right)  $ with the energy density of ordinary matter exceeding the kinetic
energy of expansion:
\[
\Omega e^{-3F_{0}}=\frac{\varkappa\varepsilon\left(  0\right)  }%
{\widetilde{\Lambda}}>1,\quad F^{\prime}\left(  0\right)  =0,\quad
\widetilde{\Lambda}>0.
\]
In the case $\widetilde{\Lambda}>0,$ $m\ll H$ the symmetric analytical
solution of the equations (\ref{Vec massive Field eq},\ref{F_,z,z = ...}) with
the initial conditions (\ref{bound cond with ord matter Lambda positive}) is
expressed in terms of trigonometric functions. The metric function $F\left(
x^{0}\right)  $, the scale factor $R\left(  x^{0}\right)  $ and the rate of
evolution $F^{\prime}\left(  x^{0}\right)  ,$
\begin{align}
F\left(  x^{0}\right)   &  =F_{0}+\frac{1}{3}\ln\left[  \left(  1-\frac{1}%
{2}\Omega e^{-3F_{0}}\right)  \cos\left(  3Hx^{0}\right)  +\frac{1}{2}\Omega
e^{-3F_{0}}\right] \label{F oscil}\\
R\left(  x^{0}\right)   &  =e^{F_{0}}\left[  \left(  1-\frac{1}{2}\Omega
e^{-3F_{0}}\right)  \cos\left(  3Hx^{0}\right)  +\frac{1}{2}\Omega e^{-3F_{0}%
}\right]  ^{\frac{1}{3}}\label{R oscil}\\
F^{\prime}\left(  x^{0}\right)   &  =H\frac{\sin\left(  3Hx^{0}\right)
}{\left(  1-\frac{2}{\Omega e^{-3F_{0}}}\right)  ^{-1}-\cos\left(
3Hx^{0}\right)  },\qquad\widetilde{\Lambda}>0,\text{ \ }m\ll H,
\label{F_,z oscil}%
\end{align}
are periodic functions with no singularity, see red curves in Figures
\ref{fig12}a,b. In the case $\widetilde{\Lambda}>0$ the origin $x^{0}=0$\ is a
point of maximum of the scale factor $R\left(  x^{0}\right)  .$ The points of
minimum (where $\cos\left(  3Hx^{0}\right)  =-1$) are
\begin{equation}
x^{0}=x_{n}^{0}=\frac{\pi}{3H}\left(  1+2n\right)  ,\text{ \ \ \ }n=0,\pm
1,\pm2,.... \label{points of minimum}%
\end{equation}
\ 

For the values of the parameters $m/H=0.02,$ $\Omega e^{-3F_{0}}=1.032$
(barely exceeding the boundary $\Omega e^{-3F_{0}}=1$) there is no difference
in Figures \ref{fig12}a,b between the curves found numerically and
analytically.
\ \ \ 

As it follows from (\ref{bound cond with ord matter Lambda positive}) without
the massive field $\left(  \phi_{0}=0\right)  $ the solutions with positive
$\widetilde{\Lambda}$ \ are possible only if the parameters are fine tuned $:$%

\begin{align}
\Omega e^{-3F_{0}}  &  =1,\label{fine tuning}\\
F^{\prime}\left(  x^{0}\right)   &  =-H\tan\frac{3Hx^{0}}{2}%
,\label{F_,z (z)  no field}\\
R\left(  x^{0}\right)   &  =e^{F_{0}}\left(  \cos^{2}\frac{3Hx^{0}}{2}\right)
^{\frac{1}{3}}. \label{R(z)  no field}%
\end{align}
The scale factor $R/R\left(  0\right)  $ (\ref{R(z) no field}) and the rate of
expansion $F^{\prime}/H$ (\ref{F_,z (z) no field})\ are the blue curves in the
Figures \ref{fig12}a and \ref{fig12}b. These fine tuned$\ $solutions have a
periodic singularity at $x^{0}=x_{n}^{0}$ (\ref{points of minimum}). In the
vicinity of each singular point $x_{n}^{0},$ as well as at $H\rightarrow0,$
the Hubble constant $H$ drops out, and the scale factor (\ref{R(z) no field})
reduces to
\begin{equation}
R\left(  x^{0}\right)  =\left(  \frac{3\varkappa\varepsilon_{0}}{4}\right)
^{1/3}\left\vert x^{0}-x_{n}^{0}\right\vert ^{2/3},\quad\left\vert \frac
{x^{0}}{x_{n}^{0}}-1\right\vert \ll1. \label{R(x^0) =   FRW}%
\end{equation}
(\ref{R(x^0) = FRW}) is the scale factor of the Friedman-Robertson-Walker
\cite{Friedman}, \cite{Robertson}, \cite{Walker} cosmology with dust matter in
the plane space geometry. The longitudinal timelike vector field$\ \phi
_{0}\neq0$ removes the singularity, see red curves in the Figures \ref{fig12}a
and \ref{fig12}b.

\paragraph{\label{Domain of regular solutions} Domain of regular oscillating
solutions}

The maximum value $F_{0}$ of the metric function (\ref{F oscil}) corresponds
to $\cos\left(  3Hx^{0}\right)  =1,$ and the minimum value $\frac{1}{3}%
\ln\left(  \Omega-e^{3F_{0}}\right)  $ -- to $\cos\left(  3Hx^{0}\right)
=-1.$ \ According to the \textquotedblleft calendar\textquotedblright%
\ (\ref{F(x^0*)=0}) today's date $x^{0\ast}$ is determined by $F\left(
x^{0\ast}\right)  =0.$ This means that $F_{0}\geqslant0,$ and $\ln\left(
\Omega-e^{3F_{0}}\right)  \leqslant0.$ Otherwise, if the today's zero value of
the metric function $F$ is outside the interval of its variation, the
oscillating solutions (\ref{F oscil}-\ref{F_,z oscil}) are not related to the
evolution of the universe. Hence, on the map of the parameters $\Omega\left(
F_{0}\right)  $ the domain of regular oscillating solutions is limited by
$F_{0}\geqslant0,$ $e^{3F_{0}}<\Omega\leqslant1+e^{3F_{0}}$. The oscillating
scenario of evolution is possible if $\Omega>1.$ One of the two: either the
estimate (\ref{Omega 0.06}) is understated, or the oscillating scenario has
nothing to do with evolution of the universe. The same applies to the
Friedman-Robertson-Walker scenario \cite{Friedman}, \cite{Robertson},
\cite{Walker}, existing on the lower boundary $\Omega=e^{3F_{0}}$
(\ref{fine tuning})$.$

The positivity of the energy $T_{00}>0$ (see (94.10) in \cite{Landau-Lifshits}%
), supported by the Lorentz gauge restriction (\ref{Lorentz gauge})
\cite{Bogolubov Shirkov}, promotes the mutual attraction between material
objects, and excludes the possibility\ of repulsion. In thirties the dark
matter had not been taken seriously. Under the action of only contracting
forces the observed expansion of the universe as a whole was considered as an
explosion of some extremely small highly compressed source. Accordingly, the
Friedman-Robertson-Walker solutions \cite{Friedman}, \cite{Robertson},
\cite{Walker} inevitably contained a singularity and existed only under\ the
fine-tuning\ restriction (\ref{fine tuning}).

The discovery of accelerated expansion confirms that in addition to the
ordinary matter there is some medium named dark matter with repulsive
properties. This source of acceleration exists for a long time after the
mysterious big bang. The longitudinal timelike vector field with a simple
Lagrangian (\ref{L dark}) turns out an appropriate tool for macroscopic
description of the dark matter, including its repulsive ability. Abandoning
the Lorentz gauge restrictions\ we get a variety of regular solutions with no
need of any fine tuning.

The idea of oscillating Universe has been proposed earlier by Lessner
\cite{Lessner} as an alternative approach to cosmology. Meanwhile, I would
better call it additional, for both kinds of regular scenarios -- cosmological
and oscillating -- are derived completely within the frames of the
Eiler-Lagrange approach and General relativity.

\section{\label{Summary}Summary}

The non-gauge vector field with the most simple Lagrangian (\ref{L dark})
turns out an appropriate tool for macroscopic description of the dark sector.
The dark substance is described via the covariant vector field equations
(\ref{Field eqs b=c=0}) and the energy-momentum tensor (\ref{Tik b=c=0}). No
longer need to invent its own model of dark matter for understanding each
observed astrophysical phenomenon.

In the galaxy scale 10 to 100 Kpc the dark matter, described by the spacelike
$\left(  \phi^{K}\phi_{K}<0\right)  $ vector field, is responsible for a
plateau in galaxy rotation curves. In the scale of the whole universe the
timelike $\left(  \phi^{K}\phi_{K}>0\right)  $\ vector field is
\textquotedblleft the missing link in the chain\textquotedblright, necessary
to understand the main features of evolution of the Universe, and avoid,
better say -- resolve, the Big Bang singularity. A mysterious Big Bang is no
longer the inevitable property of the Universe evolution, provided that the
dark matter is taken into account. The reason, why the singularity was
considered inevitable, is the impossibility of such a regular solution, where
mutually attracting objects fly away from each other.

The macroscopic description of the dark sector is applicable for studying the
structure of the Universe in the intermediate range Mpc to hundred Mpcs
\cite{Shandarin}. The field equations (\ref{Field eqs b=c=0}) and the
energy-momentum tensor (\ref{Tik b=c=0}) of dark substance allow to avoid
unnecessary model assumptions. It would be interesting to trace how additional
attraction by a\ spacelike field, dominating in the galaxy scale, transforms
into elastic repulsion of a timelike field, dominating in the scale of the
whole universe. In the intermediate range all three components of the dark
sector (described by zero-mass field, spacelike field, and timelike field)
should be taken into account altogether.

Most likely, the manifestation of dark matter in the scale of the solar system
is a fantastic, but still it is worth tracing the acceleration $\nu^{\prime
}\left(  r\right)  $ (\ref{nu'=...+lambda/r}) along the two spacecraft Pioneer
10 and Pioneer 11 hyperbolic orbits at distances between 20 - 70 AU from the
Sun. Who knows?

It is rather involuntarily, but the modern interpretations of the
observational data are mostly based on the idea of the Big Bang birth of the
Universe. The cosmic background radiation, among other phenomena, definitely
testifies that the Universe had been strongly compressed in the past. But how
strongly? The information from the past, coming to us with electromagnetic
waves, tells us only about the phenomena that happened after the Universe
became transparent. The far extrapolation to the Plank's era is based on the
assumption that the singularity is an inevitable property of cosmological
solutions of Einstein equations. The discovery of accelerated expansion
strictly pointed on the existence of hidden sector, able to resist
compression. The macroscopic theory does not restrict the strength of
compression. The degree of maximum compression is determined by the parameter
$F_{0}.$ At large negative $F_{0}$ the regular expansion after the turning
point resembles inflation. I draw attention to Fig.\ref{fig10}, where for
$F_{0}=-10$ ($R\approx1/22000$) there is a 10 order difference in the
horizontal and vertical scales. However, one should keep in mind that the dust
matter approximation is applicable if the galaxies are located at far
distances from one another.

In accordance with the Subsection \ref{Regular cosmological solutions}, the
observed \cite{Suzuki} point of minimum of $F^{\prime},$ where the
deceleration turns back to acceleration, corresponds to $m/H\sim$ $1$ (see
Figure \ref{fig11}). A timelike vector field can hardly be associated with a
massive quantum particle. Nevertheless it is worth trying to detect an
extremely light quasiparticle -- elementary excitation of some nonconventional
medium like "ghost condensate", or "aether" -- with a quantum of ground energy
$mc^{2}\sim\hbar H\sim10^{-33}eV.$

On the contrary, for a spacelike field there is a frame of reference where
$\phi_{0}=0$. The estimate for the rest energy of a particle associated with a
spacelike field with the wavelength $\lambdabar=\frac{\hslash}{mc}\sim15$ kpc
(see Figure \ref{Fig5}), is $mc^{2}\sim\allowbreak10^{-27}$ eV.

Though the macroscopic theory describes the galaxy rotation curves and various
scenarios of the universe evolution, the physical origin of dark matter and
dark energy still remains unknown. There is a hope, that applying the
energy-momentum tensor of the dark sector (\ref{Tik b=c=0}) and considering
the ordinary matter as a degenerate relativistic Fermi gas, it would be
possible not only to find the connection between the parameters of dark an
ordinary matter, but also to describe the internal structure of a heavy black
hole with no singularity in the center.


\begin{thebibliography}{99}
\bibitem {FOOTNOTE}In the derivations below the units with $c=1,$ and
$\hbar=1$ are used.
\end{thebibliography}
\end{document}